\documentclass[conference]{IEEEtran}
\IEEEoverridecommandlockouts

\usepackage{balance}
\usepackage{enumitem}
\usepackage{fancyhdr}
\usepackage{graphicx}
\usepackage{multirow}
\usepackage{rotating}
\usepackage{stfloats}
\usepackage{soul}
\usepackage{subcaption}
\usepackage{wrapfig}
\usepackage{xcolor}
\usepackage{xspace}

\usepackage{mathptmx}
\usepackage{upgreek} %
\usepackage{caption}
\usepackage{array}
\usepackage{tikz}
\usepackage{comment}
\usepackage{tcolorbox}
\usepackage[nointegrals]{wasysym}

\usepackage{cite}
\usepackage{amsmath,amssymb,amsfonts}
\usepackage{algorithmic}
\usepackage{graphicx}
\usepackage{textcomp}
\usepackage{xcolor}
\usepackage[hyphens]{url}

\usepackage[bookmarks=true,letterpaper=true,colorlinks,linkcolor=black,citecolor=blue,urlcolor=black]{hyperref}
\usepackage[capitalize,noabbrev,nameinlink]{cleveref}

\crefformat{section}{\S#2#1#3} %
\crefformat{subsection}{\S#2#1#3}
\crefformat{subsubsection}{\S#2#1#3}

\newcommand*\circled[1]{\tikz[baseline=(char.base)]{\node[shape=circle,draw,inner sep=0.5pt] (char) {#1};}}

\definecolor{iris}{rgb}{0.35, 0.31, 0.81}

\newcommand{\TheName}{\mbox{\textsc{Surge}}\xspace}
\newcommand{\simulator}{\mbox{ZSim}\xspace}
\newcommand{\salvage}{salvage\xspace}
\newcommand{\Salvage}{Salvage\xspace}
\newcommand{\salvages}{salvages\xspace}
\newcommand{\salvaged}{salvaged\xspace}
\newcommand{\salvaging}{salvaging\xspace}

\newcommand{\maxSpeedupLowLow}{$1.3\times$\xspace}
\newcommand{\maxSpeedupHighHigh}{$1.2\times$\xspace}
\newcommand{\maxSpeedupNoIO}{$1.24\times$\xspace}

\Crefname{figure}{Fig.}
{Figs.}
\crefname{figure}{Fig.}{Figs.}

\def\BibTeX{{\rm B\kern-.05em{\sc i\kern-.025em b}\kern-.08em
    T\kern-.1667em\lower.7ex\hbox{E}\kern-.125emX}}
\begin{document}
\pagestyle{plain}
\pdfpagewidth=8.5in
\pdfpageheight=11in

\pagenumbering{arabic}

\title{Pushing the Memory Bandwidth Wall with CXL-enabled Idle I/O Bandwidth Harvesting}

\author{
\IEEEauthorblockN{Divya Kiran Kadiyala}
\IEEEauthorblockA{School of Electrical and Computer Engineering\\
Georgia Institute of Technology\\
Atlanta, Georgia, USA\\
Email: dkadiyala3@gatech.edu}
\and
\IEEEauthorblockN{Alexandros Daglis}
\IEEEauthorblockA{School of Computer Science\\
Georgia Institute of Technology\\
Atlanta, Georgia, USA\\
Email: alexandros.daglis@cc.gatech.edu}
}

\maketitle

\begin{abstract}

The continual increase of cores on server-grade CPUs raises demands on memory systems, which are constrained by limited off-chip pin and data transfer rate scalability.
As a result, high-end processors typically feature lower memory bandwidth per core, at the detriment of memory-intensive workloads.
We propose alleviating this challenge by improving the utility of the CPU's limited pins.
In a typical CPU design process, the available pins are apportioned between memory and I/O traffic, each accounting for about half of the total off-chip bandwidth availability.
Consequently, unless both memory and I/O are simultaneously highly utilized, such fragmentation leads to underutilization of the valuable off-chip bandwidth resources.
An ideal architecture would offer \textit{I/O and memory bandwidth fungibility}, allowing use of the aggregate off-chip bandwidth in the form required by each workload.

In this work, we introduce \TheName, a software-supported architectural technique that boosts memory bandwidth availability by \salvaging idle I/O bandwidth resources.
\TheName leverages the capability of versatile interconnect technologies like CXL to dynamically multiplex memory and I/O traffic over the same processor interface.
We demonstrate that \TheName-enhanced architectures can accelerate memory-intensive workloads on bandwidth-constrained servers by up to {\maxSpeedupLowLow}.

\end{abstract}

\section{Introduction}
\label{sec:intro}

Modern server-grade CPUs with growing core counts are appealing to datacenter operators as enablers of improved workload consolidation and, by extension, total cost of ownership.
However, challenges in scaling memory bandwidth commensurately with cores result in reduced bandwidth-per-core availability for CPUs at the higher end of core count.
Therefore, such manycore CPUs are more vulnerable to the well-known memory bandwidth wall challenge~\cite{wulf:hitting}.

The memory wall arises from limited off-chip bandwidth: CPU pin density has only doubled every six years \cite{stanley-marbell:pinned}, while data transfer rate scaling faces power and signal integrity limits~\cite{ddr5_design_challenges,chatterjee:system}.
Because of this pin scarcity, statically partitioning off-chip bandwidth between memory and I/O at design time risks wasteful \textit{bandwidth stranding}. 
For example, a workload may saturate memory bandwidth while leaving I/O bandwidth underused (or vice versa). 
In such cases, workloads become bandwidth-bottlenecked despite unused off-chip capacity. 
With the common 1:1 memory-to-I/O provisioning ratio, the fraction of stranded bandwidth can be substantial.

Eliminating stranding requires treating all off-chip bandwidth as interchangeable between memory and I/O.
This work targets one instance of that problem: relieving memory bandwidth pressure on modern manycore CPUs.
Our key insight is that servers often leave much of their I/O bandwidth underutilized, which we propose to \textbf{s}alvage for \textbf{g}ains in \textbf{e}fficiency (\TheName).
For instance, PCIe-attached SSDs are often used only as boot volumes \cite{backblaze:ssd, narayanan:migrating}, so they see little activity during normal operation.
Similarly, many workloads generate little network traffic, leaving high-bandwidth NICs idle \cite{zhang:drack}. 
For example, Benson et al. report that (i) 70\% of server network links use less than 1\% of capacity \cite{benson:network}, and (ii) the 95th-percentile utilization over 10 days is under 25\%~\cite{benson:understanding}.
As a result, significant CPU bandwidth capacity designated for I/O at design time often remains unused across datacenter servers.
\TheName dynamically converts this idle I/O bandwidth into extra memory bandwidth when needed.

In this paper, we investigate the potential of integrating our proposed \TheName technique in server-grade systems.
To maximize \TheName's potential, our design encompasses hardware extensions to support dynamic I/O and memory traffic multiplexing on the same CPU off-chip interface, along with auxiliary software at the OS memory manager and cluster manager level.
We identify the emerging Compute Express Link (CXL) interconnect~\cite{cxl-3.2} as a timely and fitting technology to support \TheName's interface multiplexing requirement, and use it to evaluate two \TheName embodiments---\TheName Solo and \TheName Pod---that introduce different memory provisioning requirements with associated cost and utility implications.

Using \TheName to \salvage memory bandwidth over an underutilized I/O link requires provisioning a memory device (``\salvage memory'') connected via that multiplexed link.
In \TheName Solo, \salvage memory is directly connected to a single CPU.
Whenever that CPU exhibits high I/O activity, \salvage memory becomes unreachable, causing wasteful and costly \textit{memory capacity stranding.}
\TheName Pod alleviates this drawback by pooling \salvage memory within small CPU groups.
Our analysis shows \TheName Pod’s benefits for resource provisioning and performance.

We evaluate our proposed \TheName architecture using throughput workloads and online services with a range of I/O utilization scenarios.
\TheName benefits memory-bound workloads, yielding speedups of up to {\maxSpeedupNoIO} when the server's I/O subsystem is idle and up to {\maxSpeedupHighHigh} even under high I/O activity. 
In summary, we make the following contributions:
\begin{itemize}[noitemsep,topsep=2pt,leftmargin=1.1\parindent]
    \item We propose idle I/O bandwidth \salvaging to alleviate the memory wall for bandwidth-strapped manycore CPUs.
    \item We introduce \TheName, a novel software-assisted architectural mechanism that leverages  opportunistic I/O bandwidth \salvaging to boost memory bandwidth availability.
    \item We evaluate a CXL-based implementation of \TheName and analyze resource-provisioning tradeoffs introduced by two distinct architectural embodiments: Solo and Pod.
    \item We evaluate \TheName with diverse workloads and I/O activity levels, achieving speedups of up to {\maxSpeedupLowLow}.
\end{itemize}
\smallskip
\noindent \textbf{Paper outline:}
\cref{sec:motivation} provides background on memory and I/O bandwidth allocation in modern CPUs, and introduces the opportunity for dynamic memory and I/O traffic multiplexing.
\cref{sec:design} presents \TheName, a design that effectively leverages the technique of idle I/O \salvaging, and \cref{sec:impl} details a \TheName implementation using CXL as the enabling interface.
We detail our methodology in \cref{sec:method} and evaluate \TheName in \cref{sec:eval}.
Finally, \cref{sec:related} discusses related work and \cref{sec:conclusion} concludes.

\section{Background}
\label{sec:motivation}

\subsection{The Memory Bandwidth Wall}
\label{sec:background:bw_wall}

\begin{figure*}
\centering

\begin{minipage}{.61\textwidth}
  \centering
 \captionsetup[subfigure]{skip=0pt} %

  \subfloat[Mem. bandwidth availability per core.]{
        \includegraphics[width=.48\textwidth]{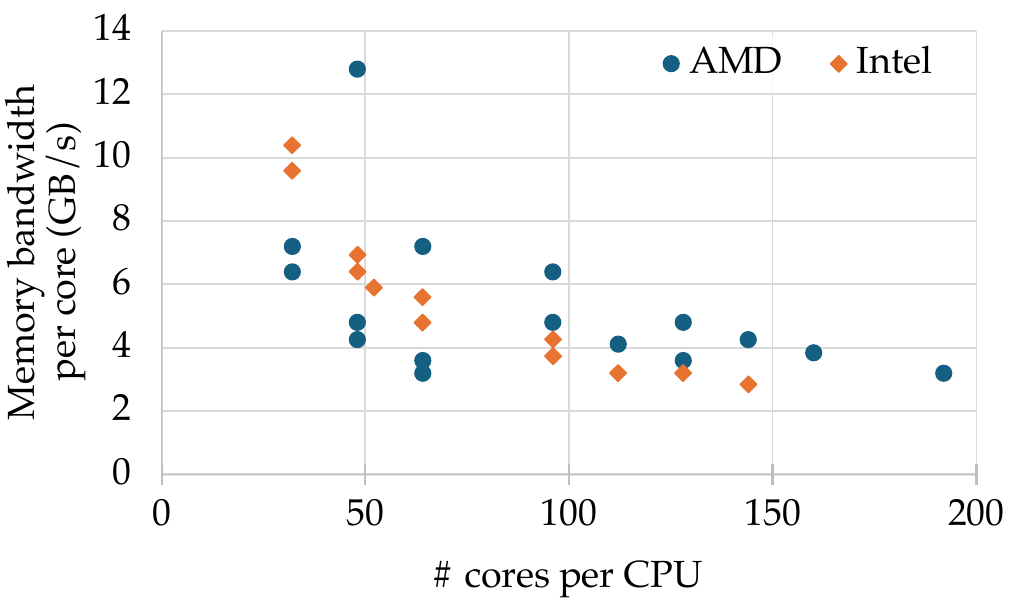}
        \label{fig:per-core-bw}
    }   
    \subfloat[Memory to I/O bandwidth availability. ]{
        \includegraphics[width=.48\textwidth]{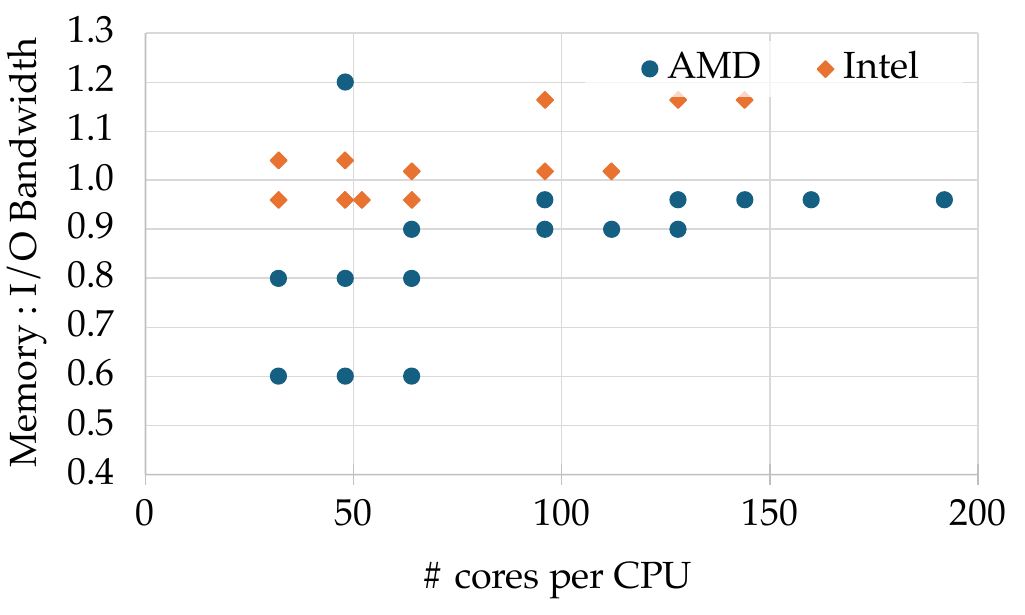}
        \label{fig:mem-to-io-bw}
    }
    \caption{Bandwidth characteristics of modern manycore processors. SKUs sampled from the AMD EPYC and Intel Xeon server processor families.}
    \label{fig:bw-availability}
    
\end{minipage}
\hfill
\begin{minipage}{.37\textwidth}
  \centering
  \includegraphics[width=\linewidth]{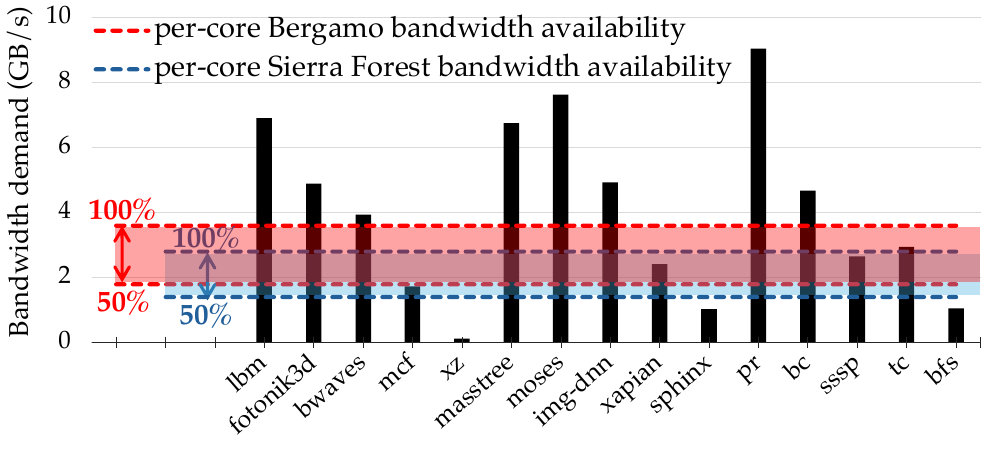}
  \captionof{figure}{Per-core memory bandwidth demands across evaluated workloads and ranges where two modern CPUs encounter queuing delays.} 
  \label{fig:workload-bw-demands}
\end{minipage}%
\end{figure*}

The memory bandwidth available per core has been decreasing in recent high-core-count processors, creating a significant performance bottleneck for bandwidth-hungry workloads. 
This trend arises from the architectural constraints of modern DDR-based memory systems. %
As core counts increase, memory bandwidth does not scale proportionally, primarily due to limitations in the number of memory channels that can be supported by a processor. 
Each DDR channel requires 288 physical pins (160+ on the processor side), which are limited by package size and signal integrity constraints \cite{cho:coaxial}. 
Furthermore, while DDR generations increase data rates, signal integrity challenges and power concerns have slowed the improvement rate  \cite{ddr5_design_challenges,zhu:package}.

\cref{fig:per-core-bw} shows the per-core memory bandwidth availability of various recent high-end manycore CPUs, as a function of core count. 
Evidently, scaling bandwidth availability to keep up with core count is challenging.
The resulting reduction in per-core bandwidth becomes especially problematic for bandwidth-intensive workloads. 
\cref{fig:workload-bw-demands} presents the per-core bandwidth requirements of several workloads we use in our evaluation (details in \cref{sec:method}). %
When compared against the per-core memory bandwidth available on processors with the highest core counts, such as Intel Sierra Forest (144 cores, 2.8 GB/s per core) and AMD Bergamo (128 cores, 3.6 GB/s per core), it is evident that many workloads approach or even exceed the bandwidth capacity available per core.
Even when the nominal bandwidth availability is not exceeded, it is well known that memory contention affects memory access time, with considerable queuing delays kicking in beyond $\sim$50\% utilization \cite{cho:coaxial, esmaili:mess,vuppalapati:tiered}.
As highlighted in \cref{fig:workload-bw-demands}, when deployed on those CPUs, the bandwidth demand of most of our evaluated workloads either exceeds bandwidth availability, or falls in the operational range of high contention.

This mismatch between workload demands and architectural supply reinforces the memory bandwidth wall as a limiting factor in the scalability and efficiency of modern manycore systems. 
In this paper, we focus on the subclass of processors at the high end of the core count spectrum to explore the implications of limited per-core memory bandwidth on workload performance and system design.

\subsection{Processor Off-chip Bandwidth Allocation}
\label{sec:background:pin_allocation}

A processor's off-chip memory bandwidth availability is capped by the limited number of pins available per socket. 
Due to physical properties and mechanical constraints, pin density scalability is challenging and lags behind silicon scaling \cite{stanley-marbell:pinned}.
The absolute off-chip data movement rate a processor can achieve over its limited pins also depends on the signaling and protocol employed---for example, serial interfaces are considerably more pin-efficient that DDR \cite{cho:coaxial}.

The available pins (and, by extension, the off-chip bandwidth they supply) are typically statically partitioned at design time between memory and I/O interfaces (typically DDR and PCIe).
Off-chip bandwidth is divided roughly equally between memory and I/O.
\cref{fig:mem-to-io-bw} shows that the memory to I/O bandwidth division ratio ranges between {0.6 and 1.2 (0.94 on average)} for the same set of server-grade CPUs shown in \cref{fig:per-core-bw}.
Such static division of off-chip bandwidth into two non-interchangeable resource groups prevents dynamic bandwidth allocation where it is most needed.
For example, a processor handling a workload that uses all available memory bandwidth but no I/O bandwidth would encounter a performance bottleneck, while leaving about half of its off-chip bandwidth resources (i.e., those reserved for I/O) idle.

\subsection{Pin Multiplexing with Compute Express Link}
\label{sec:background:cxl}

Overcoming the limitation of static, design-time off-chip bandwidth allocation by dynamically reallocating provisioned I/O bandwidth to memory can help scale past the memory bandwidth wall.
The first requirement for such an architecture is a unified protocol and interface that support seamless multiplexing of  I/O and memory traffic.

Compute Express Link (CXL)~\cite{cxl-3.2} is an emerging interconnect that provides this capability.
CXL is layered over the high-speed PCIe serial interface but replaces PCIe’s transport layer to extend functionality.
It supports three protocols: CXL.io for conventional I/O devices, CXL.mem for disaggregated memory (enabling direct ld/st access to “Type 3” memory devices), and CXL.cache for coherent accelerator attachment, which is beyond the scope of this work.
A single CXL interface supports all three protocols and can dynamically multiplex their traffic over the same physical link.
We leverage this feature to improve off-chip bandwidth utilization and mitigate the memory wall.

Owing to its underlying serial interface, CXL is highly bandwidth efficient, delivering {at least} $4\times$ higher bandwidth per pin than DDR \cite{cho:coaxial}.
However, the several SerDes conversions required on the critical path considerably increase the latency to access a CXL-attached memory device. 
This latency penalty can range from 50 to over 100 ns depending on the system architecture and distance to the memory device~\cite{li:pond, cho:coaxial}, establishing CXL-attached memory as a more likely secondary tier than a replacement for conventional DDR-attached memory.
However, while such latency premium seems prohibitive compared to the raw DRAM access latency of $\sim$50 ns, it is less so when considering the end-to-end memory access latency on a manycore CPU (e.g., 120 ns unloaded memory access time on a 128-core AMD Bergamo) and queuing delays due to contention on a highly utilized memory system, as we further analyze in \cref{sec:design:traffic_split}.

A second key differentiating CXL feature over DDR is that it is full-duplex---i.e., CXL can fully utilize its offered bandwidth in both directions simultaneously.
In contrast, DDR provides unidirectional bandwidth for memory reads and writes.
To illustrate, while a DDR5-4800 channel can move an \textit{aggregate} of 38.4 GB/s read/write traffic, a CXL-3.* x8 channel can both receive (RX) \textit{and} transmit (TX) at 32 GB/s simultaneously.
This capability highlights the opportunity for directional multiplexing of a CXL interface. 
For example, a TX-heavy workload streaming data out to an I/O device leaves plenty of bandwidth availability on the RX path that could be leveraged to boost memory read bandwidth availability.

Building on these insights, we investigate a flexible band\-width-boosting mechanism that opportunistically \salvages idle I/O bandwidth to increase memory bandwidth availability for memory-bound systems. 
Our proposed architectural approach stands in contrast to the static design-time off-chip bandwidth partitioning in current systems, enabling improved utilization of precious off-chip bandwidth resources on  modern manycore processors that dynamically adapts to workload demand.

\section{\TheName Design}
\label{sec:design}

Allowing dynamic multiplexing of memory and I/O bandwidth resources mitigates the inefficiency of statically partitioning a processor's off-chip bandwidth statically between memory and I/O at design time.
We propose \TheName, an instance of such a multiplexing approach that targets memory-bound workloads.
\TheName aims to boost memory bandwidth availability by dynamically \salvaging idle I/O resources.

\begin{figure}
    \centering
        \includegraphics[width=0.9\columnwidth]{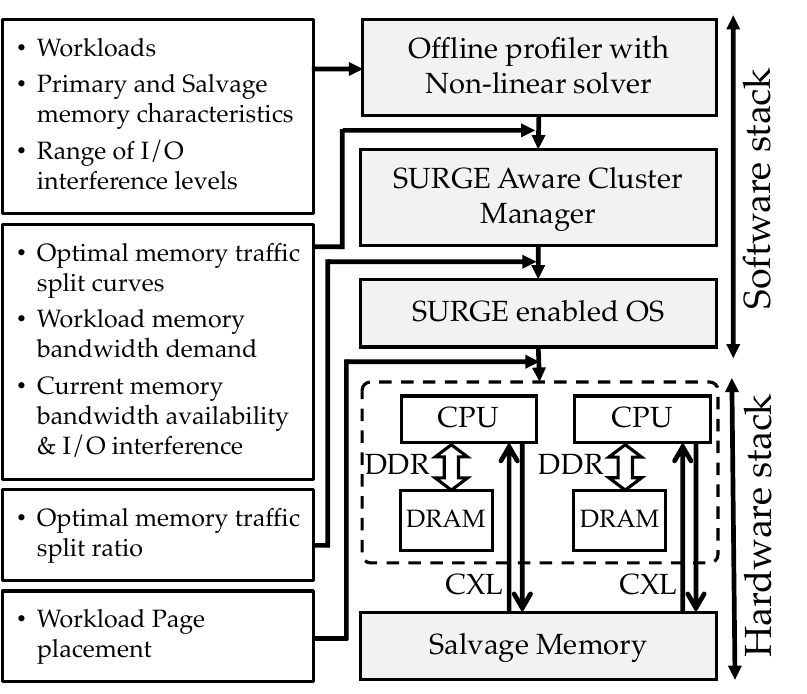}    
    \caption{\TheName hardware and software components. }
    \label{fig:overview}
\end{figure}

\cref{fig:overview} shows \TheName's high-level design.
\TheName consists of architectural extensions along with software support to maximize the hardware technique's utility.
On the architectural front (\cref{sec:design:arch}), the processor must be provisioned with additional memory that can be accessed via I/O interfaces, which must also be configured to allow multiplexing of memory traffic.
Maximizing the benefit derived from \TheName's architectural extensions requires software support that drives deployment and configuration decisions (\cref{sec:design:software}).

\subsection{Architectural Support}
\label{sec:design:arch}

We investigate two architectural design instances.
The first, simpler one targets I/O bandwidth \salvaging at the individual server level while the second extends the approach to cluster scale.
We describe the two designs in \cref{sec:design:base}--\cref{sec:design:pooled} and compare their resource efficiency in \cref{sec:design:utility}.

\TheName is not intricately tied to any specific protocol or off-chip interface fabric; it only fundamentally requires the capability of dynamic I/O and memory traffic multiplexing. 
CXL represents a promising technology meeting that requirement, thus we focus on it
to facilitate a quantitative narrative.

\subsubsection{\TheName Solo}
\label{sec:design:base}

\begin{figure}
    \subfloat[Solo \TheName. Sharing a CXL interface channel between an I/O device and a CXL Type 3 device as \salvage memory. ]{
        \includegraphics[height=5.4cm]
        {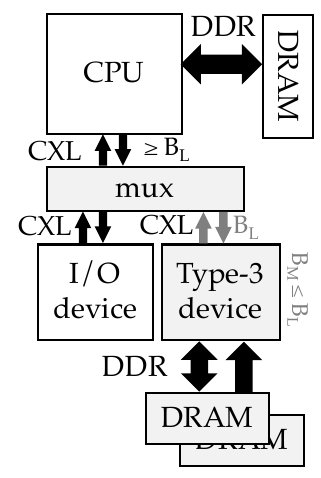}
        \label{fig:interface-sharing}
    }
    \hfill
    \subfloat[\TheName Pod. Sharing a CXL Type 3 multi-headed memory device as \salvage memory across multiple servers.]{
    \includegraphics[height=5.4cm]
    {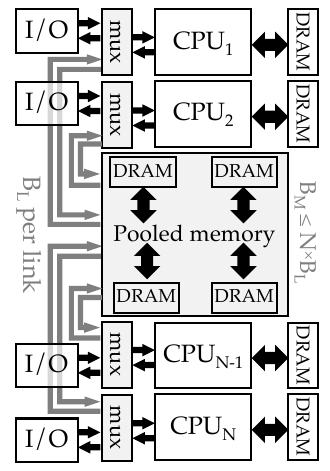}
    \label{fig:device-sharing}
    }    
    \caption{Two \TheName architectural embodiments.}
\end{figure}

\cref{fig:interface-sharing} shows the simplest form of \TheName, which targets individual servers.
Consider a CXL-capable PCIe interface with an attached I/O device, such as a NIC.
\TheName Solo bifurcates this interface: the primary link connects to the I/O device, while a secondary link connects to a memory device---in CXL terminology, a ``Type 3'' device.
We refer to the added link and memory as \textit{\salvage link} and \textit{\salvage memory}, respectively.
A simple arbiter performs the dynamic multiplexing of I/O and memory traffic, by directing each type of traffic to the respective device.
We discuss its functionality in more detail in \cref{sec:impl:hw_support}.

The hardware extension's key performance parameters are the latency overhead and bandwidth $B_L$ of the \salvage link, and the bandwidth $B_M \leq B_L$ of the \salvage memory.
The former two parameters are dictated by the used interface.
For CXL, the $B_L$ of a 16-lane PCIe 5 interface is 63 GB/s per direction, with a 50--100 ns latency overhead over direct DDR access~\cite{cho:coaxial}.
$B_M$ depends on the DDR memory channels provisioned on the Type 3 device. 
For example, a 16-lane CXL link can comfortably accommodate a DDR5-4800 channel's 38.4 GB/s of bandwidth or a couple of DDR4-3200 channels.

\begin{figure}
    \centering
        \includegraphics[width=.9\columnwidth]{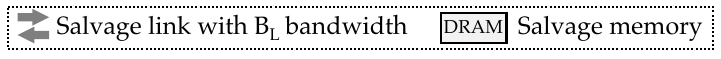}    
        \vspace{-4mm}
\end{figure}

\subsubsection{\TheName Pod}
\label{sec:design:pooled}

\TheName Pod, shown in \cref{fig:device-sharing}, is an alternative design point leveraging the same fundamental technique of opportunistic idle I/O bandwidth \salvaging, but the provisioned \salvage memory is \textit{pooled} across multiple servers that are logically grouped together in a \textit{pod}, as in prior architectures proposed to mitigate memory capacity stranding~\cite{berger:octopus,li:pond}. 
The pooled memory is a CXL Type 3 Multi-Headed Device, featuring multiple CXL ports to support direct connections to every server in the pod.
In contrast to \TheName Solo, provisioning $B_M > B_L$ is plausible, as it allows multiple servers to use the \salvage memory device simultaneously.
\cref{sec:design:utility} elaborates on $B_M$ provisioning considerations.

\subsubsection{Utility Comparison}
\label{sec:design:utility}

\begin{figure}[t]
    \centering
    \begin{subfigure}[b]{0.46\columnwidth}
        \includegraphics[width=\linewidth]{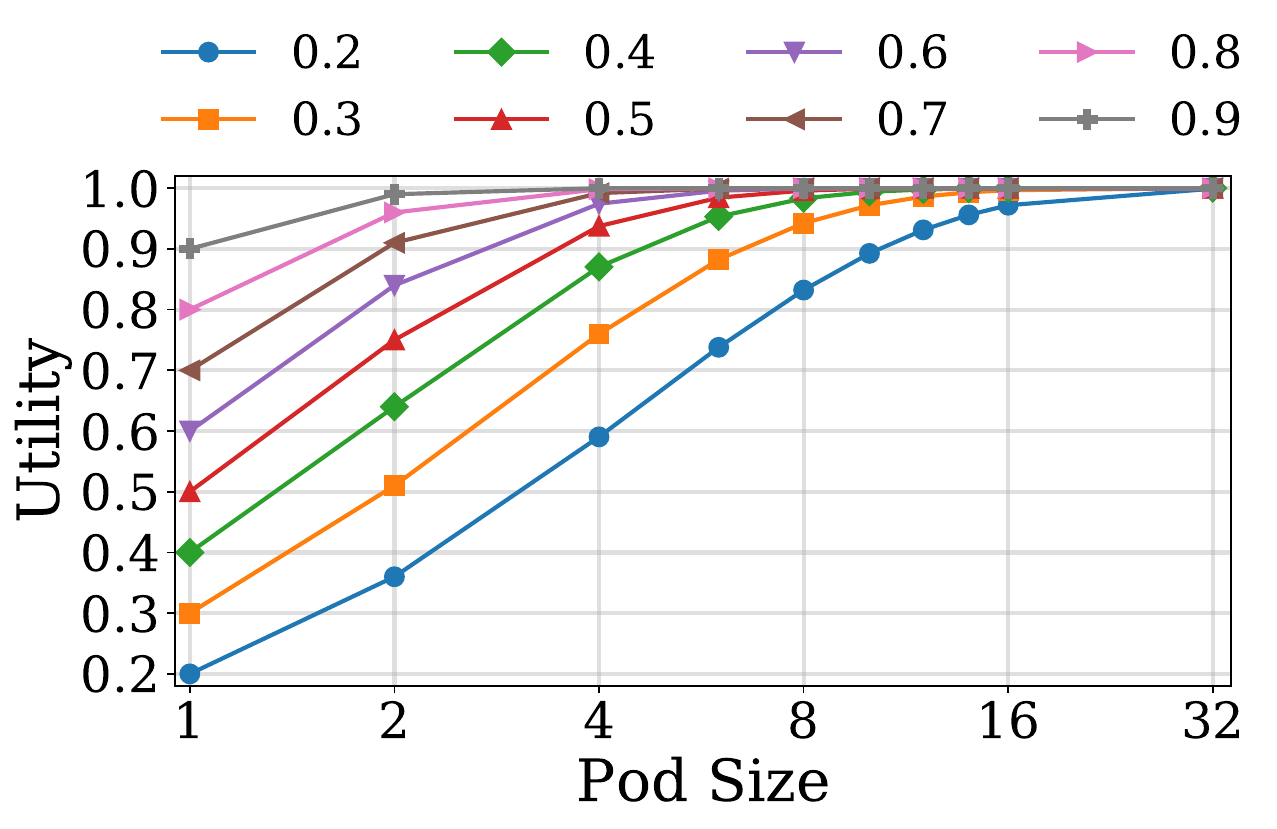}
        \caption{Utility for various P values, as a function of pod size. Pod size of 1 is \TheName Solo.}
        \label{fig:util1}
    \end{subfigure}
    \hspace{2mm} %
    \begin{subfigure}[b]{0.46\columnwidth}
        \includegraphics[width=\linewidth]{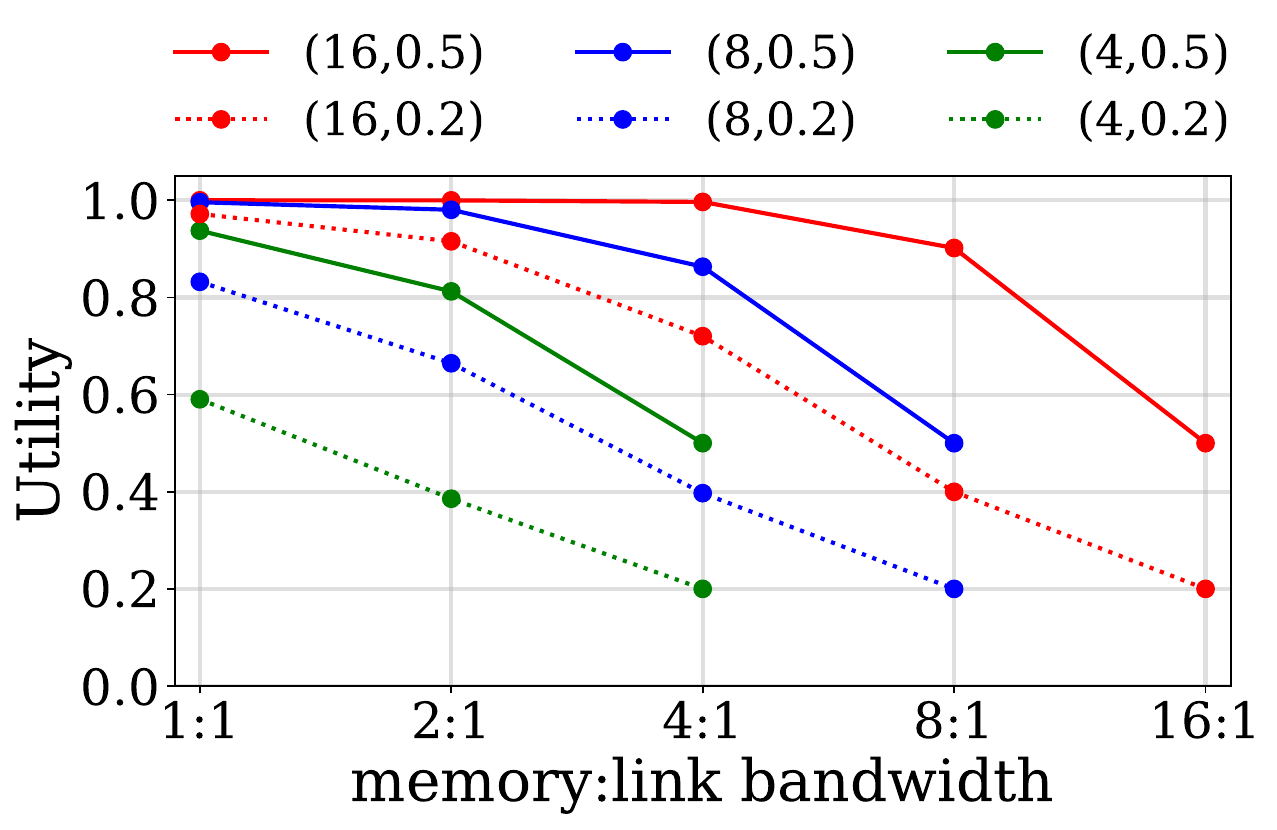}
        \caption{Utility for various (pod size, P) combinations as a function of provisioned $B_M$:$B_L$ ratio.}
        \label{fig:util2}
    \end{subfigure}
    \caption{\Salvage memory utility for \TheName as a function of pod size and P (probability of each individual \salvage link having sufficient idle bandwidth to be \salvaged by \TheName).}
    \label{fig:utility}
\end{figure}

\TheName Solo and Pod represent different design points in resource provisioning and utility.
Because \salvage memory is accessible only via a single \salvage link, high I/O activity on that link strands the memory---an undesirable and costly waste of resources.
We define \textit{utility} as the probability of using a \salvage memory’s provisioned bandwidth, with stranding probability as its inverse.
Designing a practical \TheName system thus requires balancing the cost of provisioned \salvage memory against its expected utility.
We use memory bandwidth as the primary metric, assuming it correlates directly with both memory capacity and cost.

A key drawback of \TheName Solo relative to Pod is its low utility: when a server’s I/O bandwidth is fully utilized, \salvaging memory bandwidth is impossible, leaving the attached \salvage memory stranded.
In contrast, \TheName Pod’s pooling of \salvage memory substantially reduces the probability of stranding, as  
\cref{fig:util1} demonstrates.
If the probability that a cluster's server has sufficient idle I/O bandwidth to \salvage is $P$, then the utility of \TheName Solo is $P$.
For an $N$-server pod, utility increases to $1 - \left(1-P\right)^{N}$.
For example, with $P=20\%$ and $N=16$, utility rises from 20\% to 97\%.

\TheName Pod introduces the additional consideration of memory bandwidth provisioning, $B_M$.
For an $N$-node pod with $N$ \salvage links of bandwidth $B_L$, $B_M$ may range from $B_L$ to $N \times B_L$, with utility also factoring into the decision.
\cref{fig:util1} assumed $B_M = B_L$.
\cref{fig:util2} plots utility for two pod sizes and two values of $P$ as a function of $B_M$.
On the x-axis, $X{:}1$ denotes $B_M = X \times B_L$.
Utility declines more quickly with smaller pods and higher $P$.
For example, a 16-node pod with $B_M = 4 \times B_L$ maintains over 80\% utility even with $P=20\%$.
In general, sustaining high utility favors modest $B_M$ provisioning.

While larger pods offer higher utility, they introduce two overheads: (i) greater interconnect complexity and cost, and (ii) higher latency to access remote memory, with prior work reporting 75--100 ns additional latency for groups of 8–16 servers~\cite{cho:starnuma,li:pond}.
While a cost analysis is beyond the scope of this paper, \cref{fig:utility}'s utility results indicate a pod size sweet spot around 8, with diminishing returns beyond that.
We focus on the potential of \TheName as an architectural technique.

\subsection{Software Support}
\label{sec:design:software}

\subsubsection{Workload Placement by Cluster Manager}
\label{sec:design:cluster_namanger}

\TheName's premise is that there exists ample idle I/O bandwidth to be \salvaged in order to boost memory bandwidth availability. 
Whether that condition is true depends on the workload placed on a given server.
In a cluster setting, the workload manager plays a crucial synergistic role in maximizing \TheName's applicability and utility.
Given sufficient workload knowledge and placement decisions, the cluster manager can indicate to each server's OS whether I/O bandwidth resources are expected to be underutilized. 
The OS can then use this information to initiate I/O bandwidth \salvaging, as explained in \cref{sec:design:traffic_split}.

By being \TheName-aware, the cluster manager can also avoid detrimental workload colocations. 
For example, consider workload $A$ placed on a server, which is known to minimally utilize I/O resources. 
The cluster manager decides to colocate memory-intensive workload $B$ on the same server, and notifies the server's OS that I/O bandwidth is expected to be available for \salvaging under the current workload colocation scenario.
Consequently, if workload $A$ completes, the cluster manager must either replace it with another workload with similar characteristics (i.e., one with similar I/O usage), or notify the OS about an impending change in available idle I/O bandwidth.
The OS can then adapt by migrating workload $B$'s pages from \salvage memory to the server's primary memory.

Finally, such synergy between \TheName and the cluster manager requires the latter to be aware of the memory and I/O bandwidth demands of each deployed workload.
Knowledge of workload performance characteristics is commonly required by several cluster managers, as it can greatly improve workload placement decisions, and it can be derived either via prior profiling or via explicit declaration when a workload is submitted to the cluster manager \cite{delimitrou:paragon,delimitrou:quasar,lo:heracles,verma:large-scale}.

\subsubsection{Memory Traffic Split Strategy and Analytical Model}
\label{sec:design:traffic_split}

{
Effective use of \TheName requires a judicious distribution of memory traffic between primary and \salvage memory. 
We develop an analytical model, summarized in \cref{fig:traffic_split_model}, to guide this decision.
The model's objective is to determine the traffic split that minimizes the overall Average Memory Access Time (AMAT) experienced by the workload.
As indicated in \cref{sec:design:cluster_namanger}, our key assumption is that the cluster manager has a good estimate of memory and I/O bandwidth consumption for each workload it has deployed or is about to deploy.
Given a server's current state (i.e., deployed workloads' memory and I/O bandwidth demands, memory system characteristics), our analytical model provides a method to derive the best theoretical traffic split between primary and salvage memory for the workload that is about to be deployed on the server.}

\smallskip
{\noindent \textbf{Step \circled{1}: Load-latency profiling and AMAT modeling.}
The model takes as input the profiled \textit{load–latency curves} of the server’s primary memory, \salvage memory, and salvage link, %
derived by profiling each server configuration in the cluster  \textit{once}. 
These curves provide a comprehensive view of memory system behavior under varying bandwidth utilization levels. %
Prior work demonstrated that such representation captures the performance characteristics of real memory systems well~\cite{esmaili:mess}.

For a workload with total memory bandwidth demand \( D \), a fraction 
\( R \) of the traffic is directed to the primary memory, while the remaining \((1 - R)\) is directed to the salvage memory via the \salvage link, resulting in: %
\begin{equation}
\footnotesize
    U_P = \frac{R \cdot D}{B_P}, \quad
    U_S = \frac{(1 - R) \cdot D}{B_S}
\label{eq:utilization_constraints}
\end{equation}
where \(U_P\), \(U_S\) denote the utilization levels of the primary and \salvage memories, respectively, and  \(B_P\), \(B_S\) denote their respective peak sustainable bandwidth.

The overall AMAT experienced by a workload for a given traffic split \( R \) is expressed as:
\begin{equation}
\footnotesize
\text{AMAT}(R) = R \cdot L_P(U_P) + (1 - R) \cdot \Big[
                         L_S(U_S) +
                         L_{\text{ing}}(U_{\text{ing}}) + L_{\text{egr}}(U_{\text{egr}})\Big]
\label{eq:amat_function}
\end{equation}
where \(L_P(U_p)\) and \(L_S(U_s)\) represent the
load-latency functions for the primary and salvage memories, while \(L_{\text{ing}}\) and \(L_{\text{egr}}\) capture the latency characteristics of the CXL interface's ingress and egress paths, as a function of per-direction link utilization ($U_{\text{ing}}$ and $U_{\text{egr}}$). %
$U_{\text{ing}}$ and $U_{\text{egr}}$ %
are evaluated as:
\begin{equation}
\footnotesize
U_{\text{ing}} = \frac{(1 - R)\cdot D \cdot \rho_{\mathrm{rd}}}{\eta}, \quad
U_{\text{egr}} = \frac{(1 - R)\cdot D\cdot \rho_{\mathrm{wr}}}{\eta}
\label{eq:link_load}
\end{equation}
where \(\rho_{rd} = 0.75\) and \(\rho_{wr} = 0.25\) correspond to a typical 3:1 read-to-write traffic ratio, and \(\eta = 0.94\) denotes the effective link efficiency of the CXL interface~\cite{sharma:intro-to-CXL}. Thus, as the fraction of traffic directed to the salvage memory increases, the ingress and egress link latencies also rise, capturing the queuing effects %
on the interface under higher utilization.}

\smallskip
{\noindent \textbf{Step \circled{2}: Generation of optimal traffic split curves.}
We utilize an integer linear programming (ILP) solver to determine the traffic split between primary and \salvage memory that minimizes the value of \cref{eq:amat_function}. %
Thus, the ILP solver evaluates a discrete set of candidate traffic split ratios of \( R \in [0.05,1]\) in 0.05 increments, with the optimization objective
\begin{equation}
\footnotesize
R^{*} = \arg\min_{R} \text{AMAT}(R)\text{ subject to } 0\le R \le 1    
\label{eq:traffic_split_opt}
\end{equation}
The solution \(R^{*}\) represents the theoretically \textit{optimal traffic fraction} directed to primary memory (and, correspondingly, \(1 - R^{*}\) to \salvage memory) that minimizes aggregate AMAT for a workload with bandwidth demand \(D\).
This sequence produces one of the optimal split curves shown at the end of step \circled{2} in \cref{fig:traffic_split_model}, corresponding to one system parameter combination $C = \{B_P$, $B_S$, \salvage link bandwidth ($B_{Lingr}$ and $B_{Legr}$)\}.

\begin{figure}[t]
    \centering
    \includegraphics[width=\columnwidth]{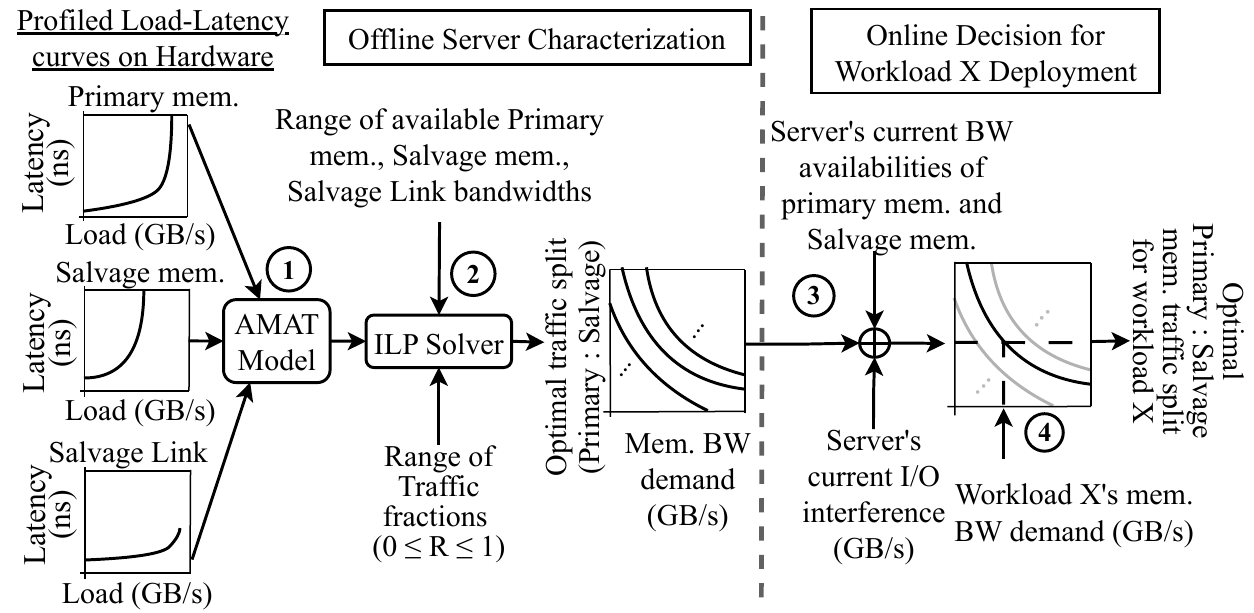} 
    \caption{Methodology to analytically determine the optimal traffic split between primary and \salvage memories.}
    \label{fig:traffic_split_model}
\end{figure}

In a typical cluster, the existence of different system configurations implies the existence of various unique system parameter combinations $C_i$.
Even for the same system, if a workload is already deployed, it already consumes bandwidth resources, implying that the remaining resource availability is a smaller $C'=\{B_P', B_S', B_{Lingr}', B_{Legr}'\}$.
Therefore, our methodology repeats step \circled{2} for a broad range of $C_i$ values, producing a large set of traffic split curves---one for each $C_i$---as shown in the middle of \cref{fig:traffic_split_model}. 
Note that this curve generation happens \textit{offline} and only \textit{once} for a given cluster.}

\smallskip
{\noindent \textbf{Steps \circled{3} and \circled{4}: Optimal traffic split selection at workload deployment. }
\label{sec:design:traffic_split_runtime}
When the cluster scheduler deploys a new workload, it determines the optimal primary-to-salvage traffic ratio for the \{workload, server\} combination under consideration, in two steps.
First (step \circled{3}), the server's current bandwidth commitments (on salvage memory, primary memory, and CXL links) due to already deployed workloads dictate the target server's resource availability $C_i$, corresponding to one of step \circled{2}'s generated traffic split curves.
Next, in step \circled{4}, the workload’s expected memory bandwidth demand \(D\) is used to probe the selected curve, yielding the traffic split $R^*$ to be applied for that workload under the current system state $C_i$. $R^*$ is conveyed to the target server's OS along with the workload deployment request.

}

\smallskip
{\noindent\textbf{Implementing the selected traffic split.} 
The desired traffic split $R^*$ can only be indirectly achieved via controlled data placement between primary and salvage memory. %
While various placement mechanisms are possible, we opt for an unintrusive page-based memory allocation approach that can be seamlessly integrated in any modern OS, as described in \cref{sec:impl:sw_support}.}

\begin{figure}
    \centering        
        \includegraphics[width=\columnwidth]{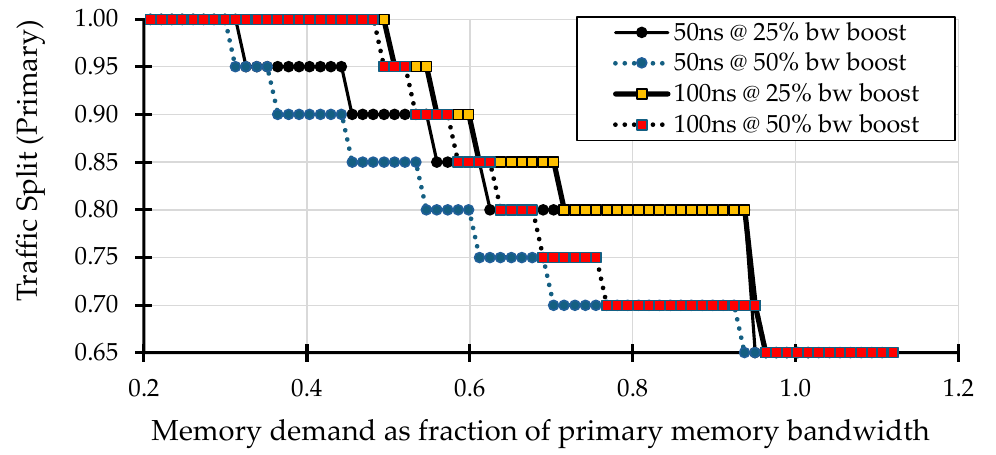}   
    \caption{Determining traffic split between primary and \salvage memory as a function of memory bandwidth demand and \salvage memory's latency/bandwidth characteristics.
    }
    \label{fig:traffic_split_demo}
\end{figure}

\smallskip\noindent\textbf{Traffic split curve example.}
\cref{fig:traffic_split_demo} exemplifies the use of the best traffic split curves.
Each of the four curves corresponds to a \salvage memory with different bandwidth and latency characteristics.
We express \salvage bandwidth relative to primary bandwidth, and latency as overhead over the raw DRAM access over DDR.
As an example, consider a system equipped with DDR5-4800 as primary memory (38.4 GB/s nominal peak bandwidth and $\sim50$ ns DRAM access latency) and \salvage memory accessed over a serial link (such as CXL), which provides a bandwidth boost expressed as fraction over primary memory bandwidth.
A \salvage memory labeled ``50 ns @ 50\% boost'' offers 19.2 GB/s at (50 + 50 =) 100 ns DRAM access latency.

The lower the memory demand, the larger the demand's fraction that should be steered to the low-latency primary memory. As demand increases, offloading more traffic to the \salvage memory reduces queuing buildup on the primary memory, reducing the overall system AMAT and improving performance. 
The effect of unloaded latency for a given memory bandwidth boost is significant at low demand levels, where queuing effects are negligible. 
At high demand levels, the system becomes bandwidth-constrained, and the contribution of queuing delays on total memory access latency starts dominating.
Therefore, under high bandwidth demand, the \salvage memory's latency overhead becomes less important, as evidenced by the four curves:
while higher \salvage bandwidth availability results in a traffic split that steers a larger fraction away from primary memory, under high bandwidth demand, the best traffic split converges to the same value regardless of the \salvage memory's latency overhead.

\section{\TheName Implementation}
\label{sec:impl}

\subsection{Hardware Extensions}
\label{sec:impl:hw_support}

As discussed in \cref{sec:design:arch}, \TheName requires a fabric that supports dynamic multiplexing of I/O and memory traffic.
Our implementation focuses on the CXL protocol, as its specification supports such dynamic multiplexing of CXL.io, .mem, and .cache traffic.
While existing CXL controller IP blocks (like the one from Rambus \cite{rambus:cxl-controller}) support all three CXL protocols, they are currently designed to operate in a single mode at system initialization time.
However, the CXL specification includes the ``Flex Bus'' feature, which supports dynamic multiplexing of traffic types and even provisions for programmable arbitration policy between them.
Flex Bus provides the basic functionality required by \TheName and we are the first, to the  best of our knowledge, to leverage this feature for new architectural designs that push the memory bandwidth wall.

\begin{figure}
    \centering
        \includegraphics[width=.7\columnwidth]{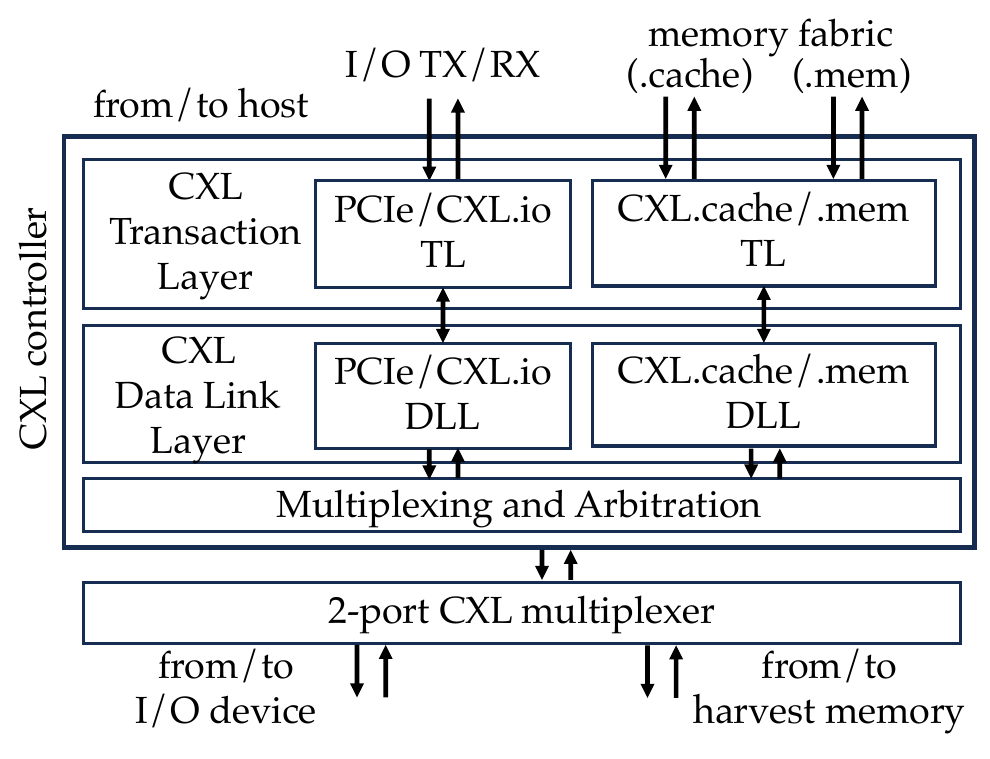}    
    \caption{I/O and memory traffic multiplexing block diagram.     }
    \label{fig:cxl-interface}
    
\end{figure}

\cref{fig:cxl-interface} shows a block diagram of a CXL controller that performs .io and .mem traffic multiplexing using a 2-port CPU-side CXL mux \cite{cxl-mux} for traffic bifurcation between the primary I/O device and our added \salvage memory device.
Because in \TheName, the interface ``belongs'' to the I/O device, we configure the arbiter to prioritize I/O over memory traffic.
At high I/O activity, such policy can penalize the---generally more latency-sensitive---memory traffic.
However, prolonged high I/O activity is beyond SURGE’s target use cases.

\subsection{Software Support}
\label{sec:impl:sw_support}

\noindent\textbf{Memory traffic split via page placement.}
\label{sec:impl:sw_support:page_placement}
In our \TheName implementation, the OS implicitly achieves memory traffic distribution between primary and \salvage memory by allocating pages in the respective memory type at the desired $R^*$ ratio.
We use a first-touch policy: every time the OS allocates a new page for a workload, it selects the primary (\salvage) memory with $R^*$ ($1-R^*$)  probability. %
{We find that such allocation results in the intended traffic split, confirming statistical expectations.}

\smallskip\noindent\textbf{Role of cluster manager.} 
We assume a cluster manager aware of the fleet's \TheName-enabled servers and at a minimum:
\begin{enumerate}[noitemsep,leftmargin=1.8\parindent,topsep=2pt]
    \item Is aware of the expected average memory and I/O bandwidth use of each workload that is about to be deployed, as derived by prior workload profiling.
    \item Given the target server's state and workload to be deployed, determines the best traffic split $R^*$ between primary and \salvage memory as per \cref{sec:design:traffic_split}'s methodology, and conveys that split to the server's OS.
    \item Does not deploy a new I/O-intensive workload on a server where a previously deployed memory-intensive workload with permission to \salvage I/O bandwidth is still active.
\end{enumerate}

\noindent Advanced software support can further improve \TheName's utility.
For example, more sophisticated page placement policies than first-touch may also explicitly take the CXL \salvage link's directionality into account and place read-/write-heavy pages in \salvage memory when the I/O interface has more TX/RX activity, respectively.
Furthermore, dynamic memory monitoring and page migration could improve adaptability over time. For instance, advanced synergies between the cluster manager and the OS can be developed, to allow dynamic memory allocation changes during a workload's runtime---the cluster manager could instruct the OS to free up I/O resources by scaling back on \salvaging via page migration from \salvage to primary memory.
We focus on the demonstration of \TheName's benefits as a novel proof-of-concept architectural technique, leaving more advanced software mechanisms to future work.

\section{Methodology}
\label{sec:method}

\noindent\textbf{Modeled system.} We simulate a scaled-down version of a modern manycore server CPU.
We consider two exemplary systems of such families.
The AMD EPYC 9754 CPU (Bergamo) \cite{wccf_bergamo} features 128 cores, 12 DDR5-4800 channels, and 128 PCIe 5 lanes per socket. 
The Intel Xeon 6780E CPU (Sierra Forest) \cite{wccf_intel_sierra} features 144 cores, 8 DDR5-6400, and 88 PCIe 5 lanes per socket. %
We model a scaled down CPU with 12 cores that preserves core-to-memory-bandwidth and core-to-LLC-capacity ratios representative of such systems.
{We do \textit{not} aim to derive an optimal pin allocation for the CPU (cf. \cref{sec:background:pin_allocation}) or argue for using a salvage memory \textit{instead} of a type-3 memory over dedicated CXL links.
\TheName's novelty lies in enabling fungible use of bandwidth availability givena pre-existing, design-time CPU pin allocation between memory and I/O, dynamically adapting aggregate resource availability to the given workload's bandwidth type demand.}

\cref{table:simSetup} details the parameters used for cycle-level simulation in ZSim \cite{sanchez:zsim}.
We adopt \TheName Pod as the primary instance of \TheName in most of our evaluation due to its higher utility, as discussed in \cref{sec:design}. 
We assume a \TheName Pod deployment offering 50\% additional memory bandwidth over a single server's primary DDR memory bandwidth at a 100 ns latency premium (at zero load).
We evaluate other bandwidth and latency values, including configurations representative of \TheName Solo deployments, as a sensitivity study (\cref{sec:eval:link-sensitivity}).

\setlength{\intextsep}{0pt}
\setlength{\columnsep}{10pt}

\begin{wrapfigure}{r}{0.46\linewidth} %
    \centering
    \includegraphics[width=0.99\linewidth]{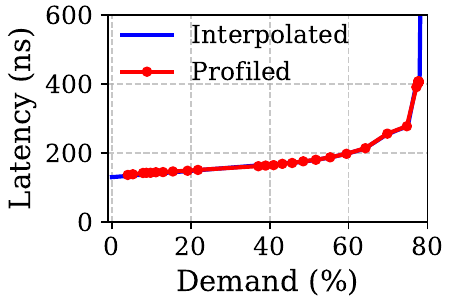}
    \caption{Profiled DDR5-4800 memory load-to-use latency vs. demand (as \% of peak) on AMD Bergamo.}
     \label{fig:mem-load-lat}
     \vspace{2mm}
\end{wrapfigure}

\smallskip\noindent\textbf{Memory modeling.} 
We model memory behavior using the Mess framework~\cite{esmaili:mess}.
We profile the memory system of several AMD and Intel server-grade CPUs and observe similar load-latency characteristics.
For our evaluation, we use the load-latency curve derived from the AMD EPYC 9754 CPU (Bergamo), as \TheName is most suitable for manycore CPUs with lower bandwidth-per-core availability.
Our simulation model uses \cref{fig:mem-load-lat}'s curve to determine the memory component's
latency as a function of its utilization, assessed at 1000-cycle execution intervals.
We apply the same modeling method for both the primary (direct DDR) and the \salvage (CXL-attached) memory.

\begin{table}[t]
\begin{center}
\begin{footnotesize}
\def\arraystretch{1.13}  
\caption{System parameters used for simulation in \simulator. 
}

\begin{tabular}{|p{1cm} | p{1.9cm}  p{2.1cm}  p{1.8cm}
|}
\hline
CPU & \multicolumn{3}{l|}{12 OoO cores, 2.4GHz, 4-wide, 256-entry ROB} \\ \hline
L1 & \multicolumn{3}{l|}{32KB L1-I\&-D, 8-way, 64B blocks, 4-cycle hit, inclusive} \\ \hline
L2 & \multicolumn{3}{l|}{512 KB, 8-way, 8-cycle hit} \\ \hline
{LLC} & \multicolumn{3}{l|}{2MB/core, shared \& inclusive, 16-way, 20-cycle hit} \\  \hline
\multirow{2}{*}{\centering Memory} & \multicolumn{3}{l|}{DDR-based (Primary): 1x DDR5-4800 channel (baseline)}\\
\cline{2-4}
 & \multicolumn{3}{l|}{CXL-attached (Salvage): 50\% bandwidth of primary} \\ \hline
{CXL} & \multicolumn{3}{l|}{1x16 PCIe 5.0, 100 ns latency overhead}  \\ \hline
{NIC} & \multicolumn{3}{l|}{1x400 Gbps interface}  \\ \hline

\end{tabular}
\label{table:simSetup}
\end{footnotesize}
\end{center}
\vspace{-5mm}
\end{table}

\smallskip\noindent\textbf{CXL interface modeling.} 
We extend \simulator's performance models with a new CXL interface component, which models the interface's latency, utilization, and the impact of the latter on the former.
We modeled an x16 CXL link operating at 32.0 GT/s, which corresponds to a 64 GB/s raw transfer rate per direction. At the Physical and Link Layers, we assume a 68-byte flit format. Each link-layer flit consists of 64B of payload and 2B of overhead, resulting in an effective link efficiency of 0.94 after accounting for protocol and formatting overheads~\cite{sharma:intro-to-CXL}.
Beyond link-layer efficiency, we also incorporate modeling of metadata transfer overheads introduced by the transport protocol for both request and response packets exchanged between the CPU host and the CXL device, following prior work~\cite{sharma:compute}.
These metadata overheads reduce the effective peak bandwidth of the CXL interface. 
For example, for a system experiencing a 2:1 read/write ratio, the effective RX (device-to-CPU) and TX (CPU-to-device) bandwidth drops to $\sim$80\% and $\sim$40\% of nominal, respectively.

\smallskip\noindent\textbf{Workloads.} %
We evaluate five workloads from each of
the  Tailbench~\cite{kasture:tailbench}, SPEC~\cite{SPEC2017}, and GAP~\cite{DBLP:beamer:GAPBS} benchmark suites.
\begin{itemize}[noitemsep,leftmargin=1.8\parindent,topsep=2pt]
    \item \textbf{Online services:} 
    From Tailbench, we evaluate an open-source search engine (xapian), an in-memory key-value store (masstree), a statistical machine translation system (moses), a speech recognition engine (sphinx), and an image recognition model (imgdnn), using their default datasets. We exclude shore, silo, and specjbb as they fail to execute reliably on \simulator.
    
    \item \textbf{Throughput workloads:} 
    From the SPEC CPU 2017 rate benchmark suite, we evaluate data compression (xz), scientific computing (bwaves, lbm, fotonik3d), and graph-based optimization (mcf) workloads in ref mode.

    \item \textbf{Graph analytics:} {From GAP, we evaluate Breadth-First Search (bfs), Single-Source Shortest Paths (sssp), PageRank (pr), Betweenness Centrality (bc), and Triangle Counting (tc),  using the Kronecker dataset.
    }
\end{itemize}

We evaluate homogeneous scenarios, where all cores of our scaled-down CPU execute the same workload.
Mixed workload scenarios yield performance approximating the average speedup achieved across single-workload experiments.
We simulate 200 million instructions per core after fast-for\-war\-ding to a representative region of interest and warming up the caches. %
\cref{fig:workload-bw-demands} shows each workload's per-core memory bandwidth demand, derived by using an ideal memory with unconstrained bandwidth and fixed zero-load latency.

\smallskip\noindent\textbf{I/O traffic modeling.} %
To evaluate I/O interference, and without loss of generality, we focus on NICs, as they are typically the highest-bandwidth I/O devices on modern servers. 
We emulate network traffic with a %
\simulator-integrated traffic generator from prior work \cite{vemmou:sweeper} that injects and ejects packets over the simulated CXL link at configurable Poisson intervals. We also capture Data Direct I/O (DDIO) behavior: all ingress network traffic is written directly into designated ways of the last-level cache (LLC). We evaluate \textit{high}, \textit{medium}, and \textit{low} NIC utilization---corresponding to 80\%, 50\%, and 10\% of peak bandwidth---capturing a range of I/O interference scenarios.

\smallskip\noindent\textbf{I/O utilization scenarios.}
We use the combination of the aforementioned workloads and network traffic modeling to investigate a range of workload deployment scenarios encountered in cloud settings. We use a \textit{$level_1$\_$level_2$} notation to indicate the level of I/O activity on the {ingress} (RX) and {egress} (TX) path, respectively, {where each level can independently take a value of the aforementioned \{\textit{low, med, high}\}. For example, \textit{low\_low} represents a case where the CPU executes workloads with little I/O activity, while \textit{low\_high} represents a scenario with a colocated media streaming workload (e.g., Netflix~\cite{netflix}), featuring heavy egress (TX) path utilization.}

\section{Evaluation}
\label{sec:eval}

We now evaluate \TheName aiming to illustrate the benefits, effectiveness, and robustness of our proposed idle I/O bandwidth salvaging technique.
We address the following questions:
\begin{enumerate}[noitemsep,leftmargin=1.8\parindent,topsep=2pt]
    \item What is \TheName's performance gain headroom in a variety of workload deployment scenarios? (\cref{sec:eval:baseline}, \cref{sec:eval:nw-interference})
    \item How effective is our heuristic in deriving the optimal traffic split for a given workload, and how robust is \TheName's performance to uncertainty in bandwidth utilization and I/O interference? (\cref{sec:eval:split-optimality}) 
    \item How much do \salvage link latency and bandwidth characteristics impact \TheName's effectiveness? (\cref{sec:eval:link-sensitivity})
\end{enumerate}

\subsection{\TheName under Low I/O Utilization}
\label{sec:eval:baseline}

\begin{figure}[t]
    \centering
        \includegraphics[width=\linewidth]{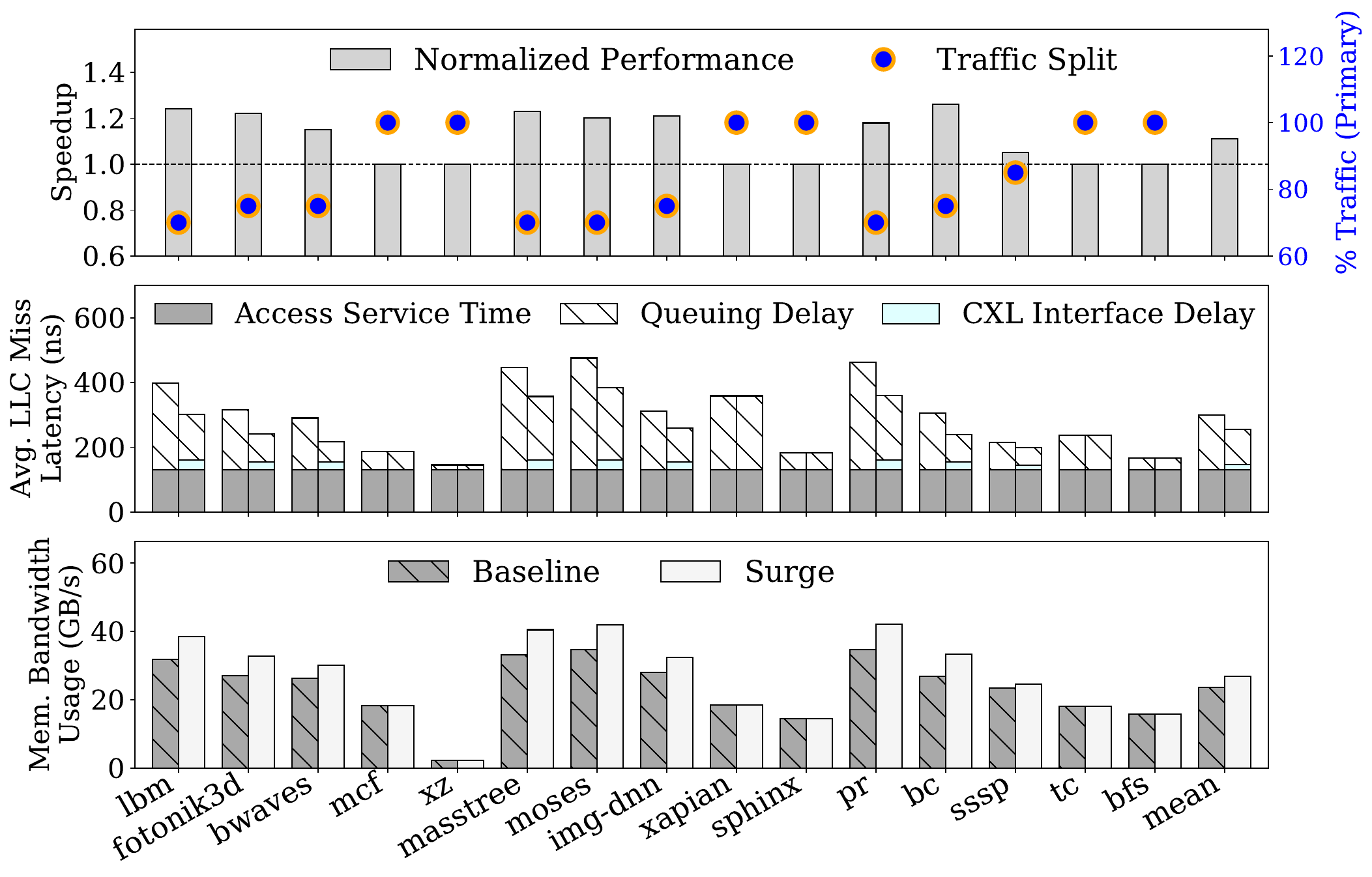}    
    \caption{Speedup, AMAT, and memory bandwidth usage for baseline vs \TheName in \textit{low\_low} networking scenario. Dots indicate the traffic split (\% of traffic to primary memory) for each workload, as determined by the \TheName methodology.    
    }
    \label{fig:baseline-speedup}
\end{figure}

 \cref{fig:baseline-speedup} (top) shows the speedup achieved by \TheName over the baseline across our evaluated workloads, under the low I/O scenario.
\cref{fig:baseline-speedup} (middle) presents the breakdown of average LLC miss latency (henceforth AMAT for brevity), decomposed into ``Access Service Time''---i.e., the unloaded latency of retrieving data from the DRAM device---and ``Queuing Delay''.
In the baseline, queuing delay occurs exclusively at the primary memory, whereas in \TheName the metric represents the average queuing latency experienced across both the primary and \salvage memories.
``CXL Interface Delay'' quantifies the time memory accesses spend on the CXL interface \textit{on average}; for example, an access to primary memory experiences zero CXL interface delay.
Finally, \cref{fig:baseline-speedup} (bottom) shows each workload's aggregate memory bandwidth demand.

Memory-intensive workloads encountering high queuing delay due to memory bandwidth pressure in the baseline are relieved by leveraging the additional memory bandwidth afforded by the \salvage memory, which is accessible over the \salvage link. 
By splitting memory traffic between the primary and \salvage memories, the system’s overall memory bandwidth increases and AMAT decreases. 
Across all workloads that encounter substantial queuing delay in the baseline, \TheName substantially reduces AMAT.
For this subset of workloads, \TheName directs {25--30\%} of the traffic to the salvage memory, resulting in an average speedup of {$1.2\times$} (up to {$1.3\times$}).
It is noteworthy that while AMAT and performance are qualitatively correlated, the same reduction in AMAT across different workloads does not deliver the same speedup. 
Individual workload characteristics, such as sensitivity to memory latency, available memory-level parallelism, and temporal burstiness in memory traffic, determine the resulting speedup.

The memory bandwidth demand of several workloads (mcf, xz, xapian, sphinx, tc, bfs) is $\leq$50\% of the primary memory's peak theoretical availability.
With such low demands, these workloads encounter minimal memory queuing in the baseline, thus \TheName's traffic split methodology directs all (or most) of the memory traffic to the primary memory.
Hence, such workloads show no or marginal benefit with \TheName.

Across all workloads and with low I/O activity, \TheName achieves an average queuing delay reduction of {$36\%$}, deliving an average speedup of {$1.1\times$}, and up to {\maxSpeedupLowLow}, over the baseline.

\subsection{\TheName with Higher I/O Utilization}
\label{sec:eval:nw-interference}

While the low I/O scenarios are the most intuitive use case for idle I/O bandwidth salvaging, \TheName is applicable and effective even with higher I/O activity. 
To highlight \TheName's breadth of applicability, we evaluate scenarios with a variety of I/O utilization levels, as outlined in \cref{sec:method}. 
We use different RX\_TX traffic level combinations to evaluate \textit{low\_high, high\_low, med\_med, }and \textit{high\_high} I/O traffic interference.

\begin{figure}
    \centering
     \captionsetup[subfigure]{skip=0pt} %
    \subfloat[Baseline workload sensitivity to I/O utilization. Speedup values normalized to performance with \textit{low\_low} utilization. %
    ]{
    \hspace{-11.1pt}
        \includegraphics[width=0.93\linewidth]{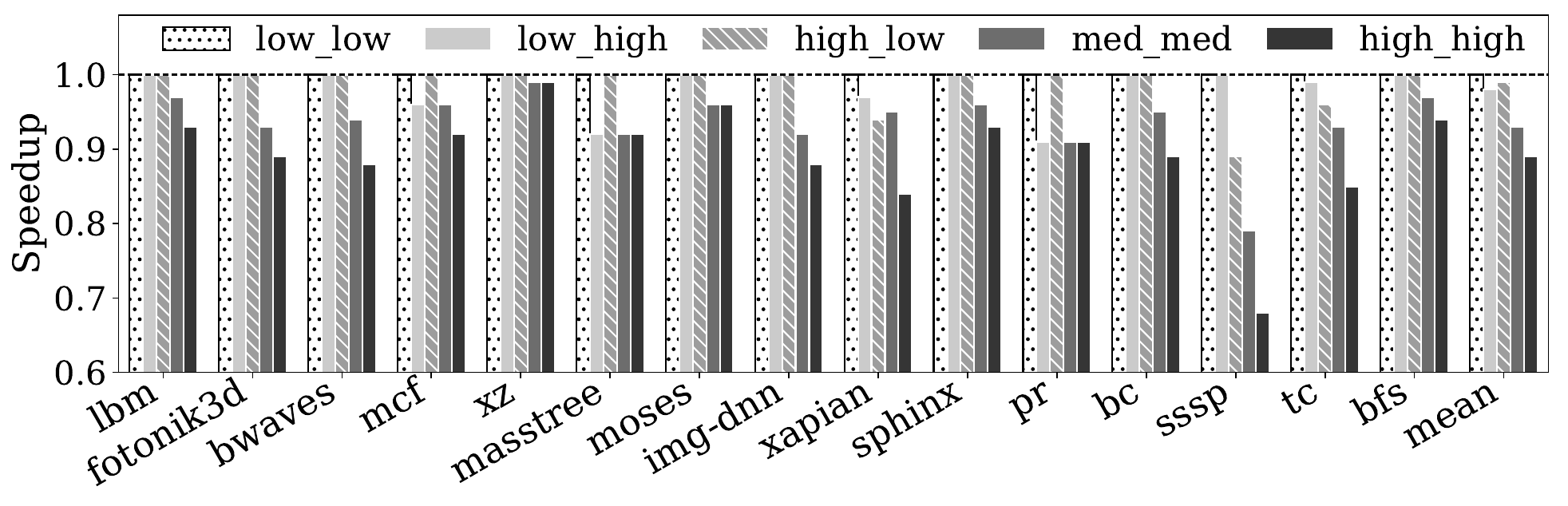}
        \label{fig:network-interference-baseline}
    }
    \vspace{2mm}
    \subfloat[\TheName benefit under different levels of I/O utilization. ]{
        \includegraphics[width=\linewidth]{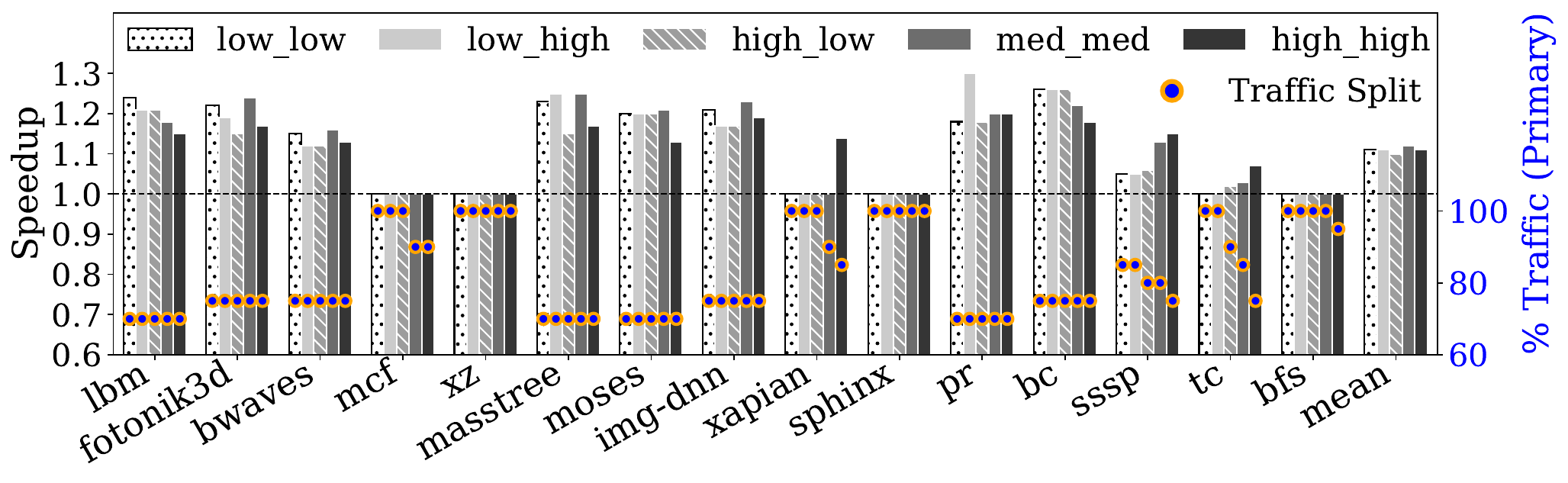}
        \label{fig:network-interference-surge}
    }
    \caption{Impact of I/O traffic interference. \textit{$level_1\_level_2$} on the legends indicates RX\_TX path utilization.
    }
\end{figure}

We first focus on the baseline to study each workload's behavior with increasing I/O traffic.
\cref{fig:network-interference-baseline} shows the results.
Even though we use DDIO, increasing I/O traffic exacerbates LLC and memory bandwidth pressure, as both application and I/O data move between the LLC and the primary memory.
{Such network-induced traffic introduces interference in the memory hierarchy that degrades performance.}

We make two observations.
First, workloads exhibit different sensitivity to increasing I/O traffic interference.
{For example, xz, which has the lowest bandwidth requirements, remains virtually unaffected, xapian, sssp, and tc exhibit sharp performance drops as I/O utilization increases, and the remaining workloads experience different degrees of more gradual degradation.
Second, workloads are affected differently by RX and TX interference, with tc and sssp more sensitive to the former, and others (like mcf, masstree, pr)  more sensitive to the latter. 
Across workloads, a medium or high interference on both RX and TX is detrimental to performance.}

~\cref{fig:network-interference-surge} shows the impact of I/O utilization on  \TheName's effectiveness.
The \textit{low\_low} configuration represents the scenario studied in \cref{sec:eval:baseline}, \textit{low\_high} represents TX-heavy scenarios, such as media streaming, while \textit{high\_low} represents RX-heavy scenarios.
\TheName yields a stable speedup over the baseline, across all I/O utilization levels, of {$\sim1.1\times$} on average.
The results indicate that \TheName effectively adjusts the traffic split between primary and \salvage memory, balancing bandwidth pressure on primary memory with I/O interface and salvage memory bandwidth availability to minimize the contribution of queuing delays on AMAT.

As I/O activity increases, \TheName's methodology reevaluates the optimal traffic split between primary and \salvage memory.
Higher I/O activity means less available bandwidth to salvage on the I/O link, but also increased bandwidth pressure on the primary memory due to network induced memory traffic, pushing queuing delays higher.
With few exceptions, the \TheName-determined optimal traffic split remains stable across I/O utilization level. This is because (i) even under higher I/O utilization, the I/O interface offers considerable bandwidth availability to complement the primary memory's bandwidth, and (ii) the increase in the I/O interface's contributed queuing delay at higher utilization is smaller than the primary memory's high queuing delays under high memory bandwidth demand.
Even in the few cases where higher I/O interference shifts the traffic split (mcf, xapian, sssp, tc, bfs), \TheName places \textit{less} traffic to primary memory, indicating that queuing at the primary memory bandwidth is more detrimental than queuing on the I/O interface, thereby determining that shifting more traffic to the latter is overall beneficial.
As a result, even with \textit{high\_high} I/O activity, \TheName yields speedups of up to {\maxSpeedupHighHigh.}
Importantly, even when \TheName does not benefit a workload, it has a performance-neutral effect, delivering performance on par with the baseline.

Finally, in a few workload cases, like xapian, sssp, and tc, \TheName delivers a higher speedup under higher I/O utilization.
The increased speedup is because the baseline's performance degrades fast, while \TheName offers better performance robustness, as its enforced traffic split manages to absorb the detrimental bandwidth interference experienced by the baseline at the primary memory.

\subsection{Robustness of the \TheName Traffic Split Methodology}
\label{sec:eval:split-optimality}

\begin{figure}[t]
    \centering
    \includegraphics[width=\linewidth]{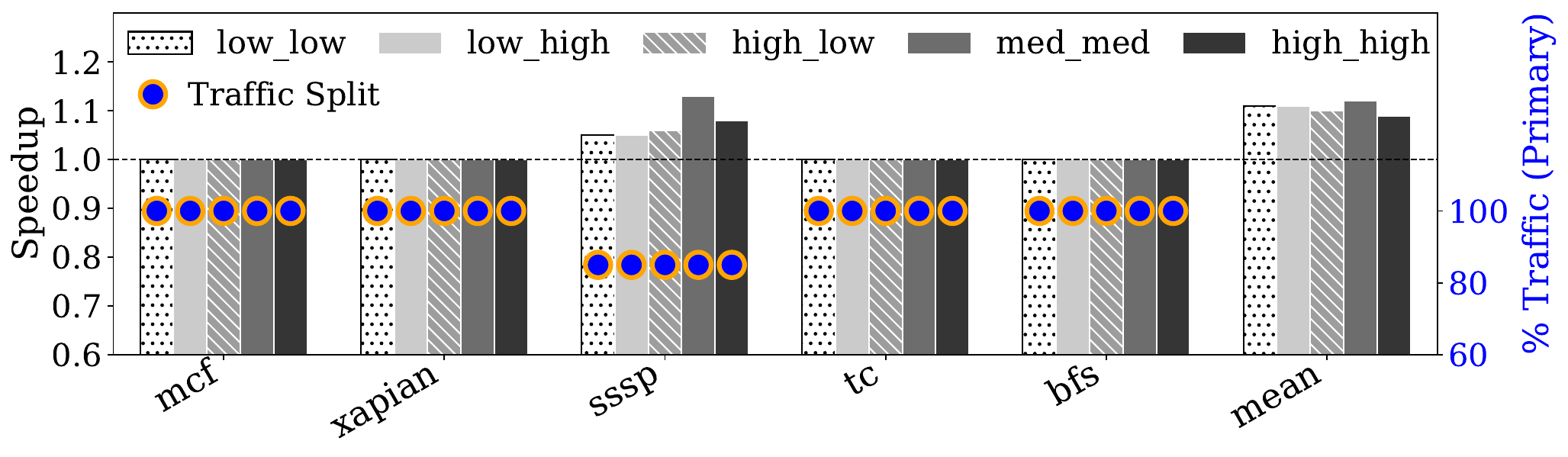}  
    \caption{Robustness to I/O interference. Traffic split configured expecting $low\_low$ I/O traffic in all cases. The plot shows only the subset of workloads that are sensitive to the \TheName-selected traffic split, while the mean refers to all 15 workloads.
    }
    \label{fig:io-robustness}
\end{figure}

\TheName's traffic split methodology relies on workload knowledge, specifically their memory bandwidth demand and I/O activity levels, derived from profiling.
In this section, we study our traffic split approach's robustness to imprecision in such knowledge, due to inaccurate profiling or fluctuations in memory bandwidth utilization and I/O interference.

\subsubsection{Performance sensitivity to unexpected I/O interference} 

We first focus on the impact of I/O interference uncertainty; namely, when I/O utilization ends up being higher than expected at the time of a workload's deployment---for example, expecting \textit{low} utilization but encountering \textit{medium} or \textit{high} instead, which represents a substantial spike of \textbf{5$\times$/8$\times$}, respectively.
\cref{fig:network-interference-surge} already showed that the optimal traffic split for a given workload is largely insensitive to the I/O utilization level.
In other words, for most workloads, the \TheName traffic split methodology would determine the same traffic split value across all the I/O utilization levels we consider.

\cref{fig:io-robustness} focuses on \TheName's performance for the five workloads that benefit from adjusting the traffic split according to the I/O activity level.
Performance for the remaining ten workloads is not shown, because it matches \cref{fig:network-interference-surge}'s results.
Of those five workloads, sssp still preserves some performance gains with \TheName.
While the remaining four lose any performance gains when I/O utilization is higher than the expected \textit{low\_low}, \textit{none} of them experiences a slowdown, performing on par with the baseline.
For some cases, like mcf and xapian under \textit{med\_med} and mcf under \textit{high\_high}, a comparison between \cref{fig:network-interference-surge} and \cref{fig:io-robustness} reveals that \TheName performs on par with the baseline, despite a difference in the traffic split used in the two experiments. 
This behavior is attributed to our observation that the optimal traffic split is a \textit{range} rather than a single point, as explained next in \cref{sec:split-sensitivity} and \cref{fig:amat-ipc}.

Overall, \TheName delivers robust performance results even under unexpected I/O traffic interference (whether sustained or instantaneous in the form of temporal spikes). 
No workload experiences a slowdown, average speedup remains substantial at {$\sim1.1\times$} across I/O interference levels, with some workloads still experiencing a performance gain of up to {$1.25\times$}.

\subsubsection{Performance sensitivity to traffic split}
\label{sec:split-sensitivity}
\begin{figure}[t]
    \centering
    \includegraphics[width=0.99\linewidth]{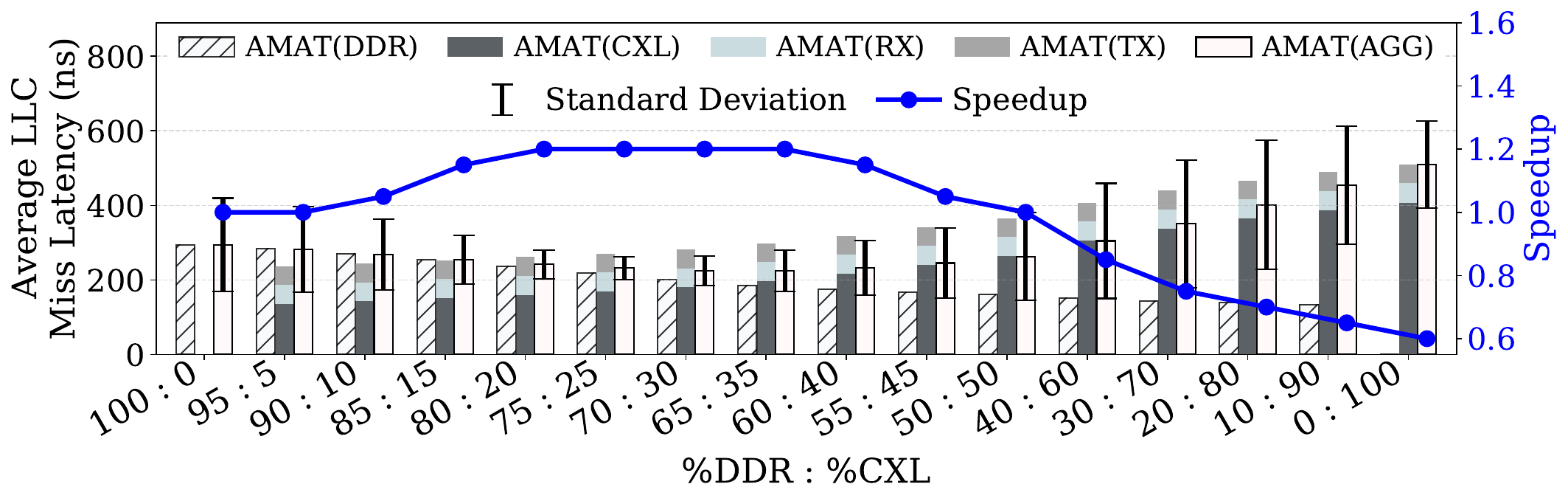}    
    \caption{AMAT and speedup for the {bc} workload across the range of possible primary:\salvage memory traffic splits. Whiskers show standard deviation.
    }
    \label{fig:amat-ipc}
\end{figure}

For our next robustness analysis, we use the workloads \textit{without any I/O interference} to exclusively focus on memory bandwidth utilization uncertainty.  
\cref{fig:amat-ipc} shows the resulting AMAT and speedup as a function of the applied traffic split.
In the interest of space, we only indicatively show data for the {bc} workload; similar trends hold for all workloads.
Evidently, the selected traffic split critically affects performance: the best choice can yield a speedup of {$1.2\times$}, but a poor choice can incur a {67\%} slowdown.
For {bc}, the optimal primary:\salvage split in a deployment scenario without I/O interference falls in the {80:20 to 65:35} range.
However, there is a wide traffic split \textit{range} from {85:15 to 60:40} 
where \TheName yields substantial performance gains of {$1.15\times$} or higher.
Given the breadth of this range, \TheName is robust to uncertainty or fluctuations in memory bandwidth usage.

\TheName's performance gains primarily stem from reducing the memory system's AMAT.
\cref{fig:amat-ipc} confirms that the highest speedup is achieved by the configuration with the lowest AMAT.
The whiskers on \cref{fig:amat-ipc}'s aggregate AMAT bars highlight an additional significant benefit of \TheName.  
In addition to reducing AMAT, \TheName also reduces memory access latency variance, because queuing delay---which significantly fluctuates over time due to temporal bursts in memory bandwidth demand---becomes a smaller contributor to overall latency.
Higher memory access time predictability additionally contributes to performance improvement, as observed in prior work
\cite{coaxial_arxiv}.

\TheName's methodology determines a traffic split based on average bandwidth utilization derived from profiling. 
We now evaluate how close the derived traffic split is to the optimal (i.e., best-performing split).
We derive the latter experimentally, {using a primary:\salvage split sweep in 5\% increments.}

\begin{figure}[t]
    \centering
        \includegraphics[width=0.99\linewidth]{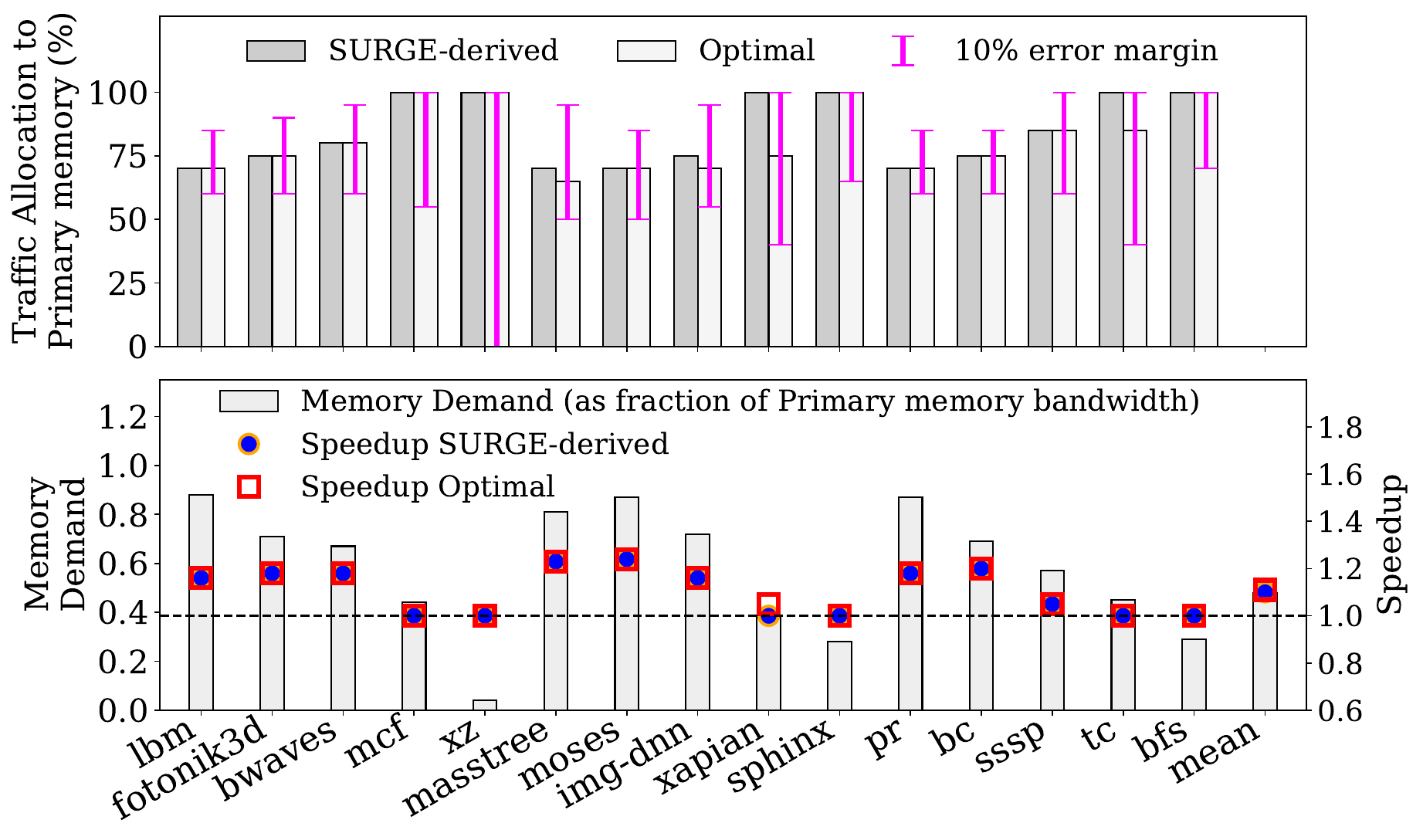}
    \caption{Deviation of \TheName-derived traffic split from optimal split, and impact on performance in \textit{no I/O} scenarios.}
    \label{fig:prediction-error}
\end{figure}

\cref{fig:prediction-error} (top) shows that, across all workloads, the \TheName methodology's theoretically derived traffic split matches or is very close to the practically optimal.
Importantly, there is a significant margin of deviation from the optimal split within which \TheName performs close to its maximum potential: the error bars indicate the traffic split range that still delivers performance within {10\%} of the optimal split.
The reason for such robustness is that there is a wide plateau around the performance peak, as previously demonstrated in \cref{fig:amat-ipc}.

\cref{fig:prediction-error} (bottom) shows each workload's memory demand, and the speedup achieved with the \TheName-derived and optimal split, respectively.
Indeed, \TheName's achieved speedup is very close to the one achieved with an optimal split: within {1\%} on average and at most within {5\%}.

\subsection{Sensitivity to \Salvage Link Latency and Bandwidth}
\label{sec:eval:link-sensitivity}

\begin{figure}[t]
    \centering
    \includegraphics[width=0.995\linewidth]{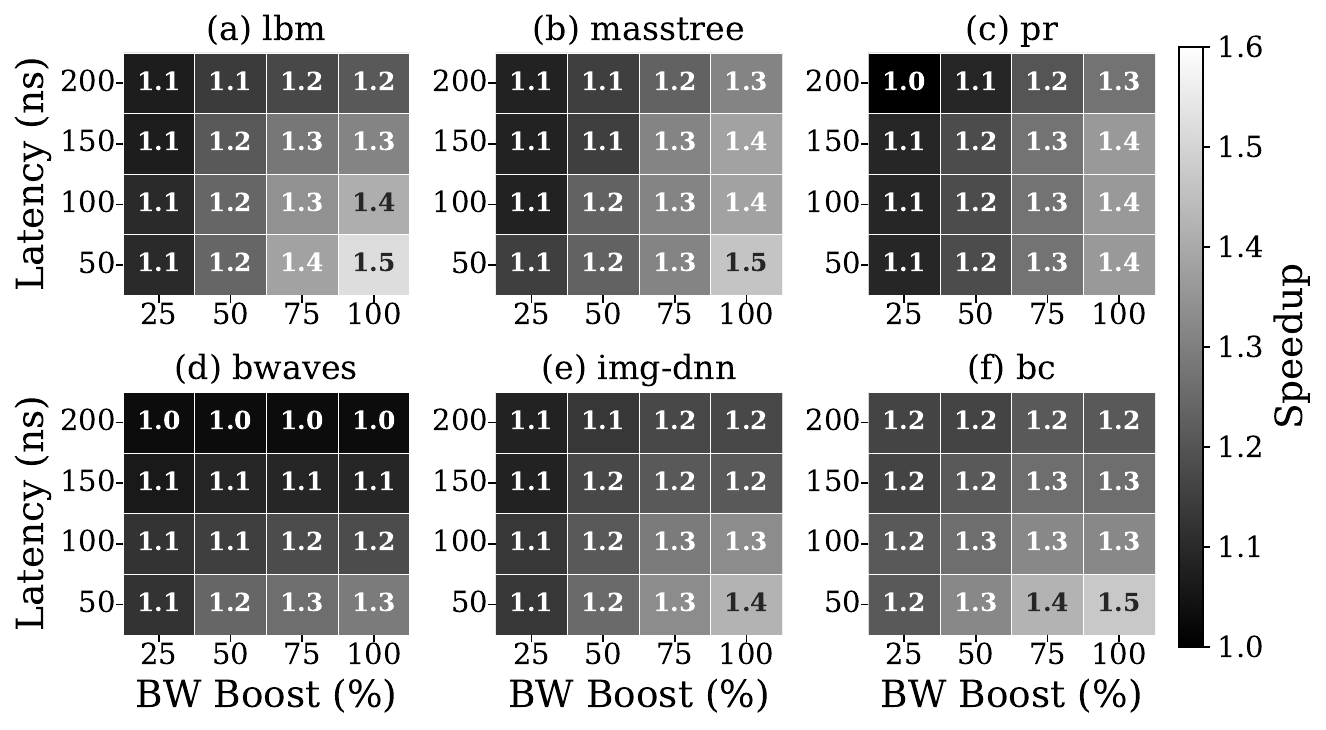}
    \caption{Impact of \salvage memory's bandwidth-latency characteristics on \TheName's effectiveness. \TheName Solo/Pod deployments are more likely to offer characteristics landing toward the bottom left/top right corner, respectively.
    }
    \label{fig:latency-bandwidth}
\end{figure}

We conclude our evaluation with a sensitivity study on the latency and bandwidth characteristics of the CXL-attached \salvage memory.
We express \salvage memory bandwidth in terms relative to the primary memory's bandwidth.
\cref{fig:latency-bandwidth} shows \TheName's attainable speedup for 16 bandwidth-latency combinations and six of our workloads with \textit{low\_low} I/O utilization.
50\% bandwidth boost at 100 ns latency premium is the configuration used throughout the evaluation, which we consider a reasonable design point for a \TheName Pod deployment.
Configurations with lower bandwidth boost and potentially lower latency (e.g., 25\% boost, 50 ns) are more likely for \TheName Solo deployments, as discussed in \cref{sec:design:utility}.

While the sensitivity of workloads to the \salvage memory's bandwidth-latency characteristics differs, the same key observation holds across the board: for the bandwidth-strapped scenarios \TheName targets, bandwidth is more important than unloaded latency penalty.
The reason is that \TheName is a technique that shines under bandwidth scarcity, where the primary memory experiences latency spikes due to queuing.
In such cases, even a 200 ns unloaded latency premium is preferable, if the memory system can be brought to an operational point prior to the region of exponential queuing delays.
Workloads with lower bandwidth demands that don't benefit from \TheName, like mcf (not shown), are insensitive to \salvage memory latency premium, because the \TheName traffic split methodology diverts less (or no) traffic to the \salvage memory.
{The effect of latency is more pronounced at higher bandwidth availability, where the contribution of queuing to AMAT is less pronounced.}
Our sensitivity study highlights that the higher bandwidth that \TheName Pod is likely to be able to afford can more than offset is higher latency premium.

\section{Related Work}
\label{sec:related}
High-end processors feature increasing core counts for throughput gains, with the aim of improving resource consolidation and TCO.
However, commensurately scaling memory bandwidth availability using DDR technology alone is challenging, resulting in reducing the memory bandwidth availability per core.
Hence, such manycore processors cannot be effectively utilized when handling memory-intensive workloads.
Some recent high-end CPUs alleviate the memory bandwidth wall by deploying HBM \cite{kim:hbm,wccf_intel_hbm} or large die-stacked SRAM~\cite{vcache}. However, in addition to their high cost, such exotic solutions suffer from severely limited capacity~\cite{qureshi:fundamental}.

\smallskip\noindent\textbf{Memory tiering.}
In the context of CXL, recent works have explored using CXL-attached memory as a secondary tier to expand both capacity and bandwidth~\cite{marouf:tpp,vuppalapati:tiered, arif:CXL:ICPP22, huang:cxl:in-memory-processing}. However, CXL-based tiering requires careful hardware-software co-design to mitigate the high serial link latency~\cite{Sim:CXL:memory-intensive:hpca24, aguilera:mem-disagg:SIGOPS23}. 
As in \TheName, careful page placement across the memory tiers plays an important role in maximizing performance.

\smallskip\noindent\textbf{Memory bandwidth boosting.} 
BATMAN \cite{chou:batman} proposes spreading a workload's dataset across planar and die-stacked DRAM to maximize aggregate memory bandwidth availability.
Caption~\cite{sun:demystifying-cxl:MICRO23} aims to improve the performance of bandwidth-intensive applications by leveraging CXL memory devices for bandwidth expansion, and uses a linear Machine Learning model to perform bandwidth-aware page allocation across the memory tiers. %
Sehgal et al.~\cite{Sehgal:Micron:memory-bandwidth} propose software-based weighted interleaving of pages to leverage the bandwidth of DDR and CXL memory, considering the workload characteristics and performance characteristics of memories to various memory read/write patterns. %
Coaxial proposes complete replacement of the prevalent DDR-based memory system with a bandwidth-rich CXL-based memory system~\cite{cho:coaxial}, but such radical approach can hurt performance under low system utilization.
In contrast to all prior work deploying and managing a secondary memory tier to boost bandwidth, \TheName multiplexes existing (primarily I/O) interfaces to opportunistically boost memory bandwidth availability.
\TheName is the first extensive study of the opportunities arising from the CXL-enabled fungibility of I/O and memory traffic, and the first system to leverage that capability to dynamically \salvage idle I/O bandwidth, convert it to additional memory bandwidth for memory-bound workload acceleration.

\smallskip\noindent\textbf{Page placement.} \TheName leverages controlled page placement as a way to split traffic between the tiered primary-\salvage memory system.
We found weighted first-touch page placement sufficient for the simple proof-of-concept deployment scenarios we evaluated.
However, more complex cases with workload churn would require periodic re-evaluation of each workload's traffic split and support for more sophisticated page placement and migration techniques.
There are several such mechanisms in the literature, both software-based~\cite{marouf:tpp, agarwal:thermostat:ASPLOS17, povoas:ICPE25, liu:hotness:osdi25, xiang:Nomad:osdi24, raybuck:HeMem:SOSP21} and hardware-assisted \cite{cho:starnuma, petrucci:micron:arxiv}.

\section{Conclusion}
\label{sec:conclusion}

Modern manycore CPUs encounter increasing memory bandwidth pressure, as off-chip bandwidth scales slower than core counts.
This paper identifies an opportunity to ameliorate this pressure by introducing \TheName, a technique that dynamically \salvages idle I/O bandwidth and converts it to a memory bandwidth boost.
\TheName realizes such bandwidth \salvaging by leveraging the modern CXL interface's capability for dynamic multiplexing of I/O and memory traffic.
We introduce a design that combines \TheName's architectural capability with software support that determines the best traffic split between primary and salvage memory, with the goal of minimizing each workload's AMAT.
Our evaluation shows that \TheName delivers performance gains of up to {\maxSpeedupLowLow} when the I/O subsystem is largely underutilized, and up to {\maxSpeedupHighHigh} under substantial I/O activity. 
\TheName represents a novel, effective, and timely technique to boost memory bandwidth availability for modern bandwidth-strapped manycore processors.

\section*{Acknowledgments}

We thank Marina Vemmou for extensive help with \simulator, particularly with the integration of NIC traffic and DDIO behavior; Hamed Seyedroudbari for insightful discussions; Albert Cho and Prachatos Mitra for feedback that improved the paper; and Balaadithya Uppalapati for discussions that helped firm up the ILP solver component of \TheName's traffic split methodology.
This work was supported by the National Science Foundation, award SHF-2333049.

\bibliographystyle{IEEEtranS}
\balance
\bibliography{gen-abbrev,refs}

\end{document}